\documentclass{biometrika}
\usepackage{amssymb,
	amsmath,mathrsfs,
bm,color}
\usepackage{lineno}
\usepackage{graphicx}
\usepackage{multirow}
\usepackage{epsfig}
\usepackage{natbib}
\usepackage[shortlabels]{enumitem}
\usepackage{threeparttable}

\usepackage{url}

\numberwithin{equation}{section}

\newcommand{\defn}{\stackrel{\mbox{{\tiny def}}}{=}}
\newcommand{\trans}{^{\mbox{\tiny{T}}}}

\def\I{\mathrm{I}}

\def\t{{\bf t}}
\def\x{{x}}

\def\calN{\mathcal{N}}

\def\wh{\widehat}

\def\beqr{\begin{eqnarray}}
\def\eeqr{\end{eqnarray}}
\def\beqrs{\begin{eqnarray*}}
\def\eeqrs{\end{eqnarray*}}

\def\bGam{\Gamma}

\def\bSig{\Sigma}

\def\mR{\mathbb{R}}

\newcommand{\X}{X}

\newcommand{\diag}{\mathrm{diag}}

\newcommand{\bTh}{{\Theta}}
\newcommand{\btheta}{{\theta}}

\newcommand\independent{\protect\mathpalette{\protect\independenT}{\perp}}
\def\independenT#1#2{\mathrel{\rlap{$#1#2$}\mkern2mu{#1#2}}}

\def\n{\nonumber}

\usepackage{url}
\DeclareFontFamily{U}{mathx}{\hyphenchar\font45}
\DeclareFontShape{U}{mathx}{m}{n}{
	<5> <6> <7> <8> <9> <10>
	<10.95> <12> <14.4> <17.28> <20.74> <24.88>
	mathx10
}{}
\DeclareSymbolFont{mathx}{U}{mathx}{m}{n}
\DeclareFontSubstitution{U}{mathx}{m}{n}
\DeclareMathAccent{\widecheck}{0}{mathx}{"71}
\DeclareMathAccent{\wideparen}{0}{mathx}{"75}

\newcommand{\bbeta}{\zeta}

\newcommand{\abs}[1]{\vert#1\vert}

\def\pr{\mathrm{pr}}

\usepackage{caption}

\usepackage{tikz}
\usetikzlibrary{positioning}
\tikzset{> = stealth,
    hidden/.style = {
        draw = black,
        shape = circle,
        inner sep = 1pt
    }
}

\newcommand{\QNDE}{\mathrm{qNDE}}

\newcommand{\QNIE}{\mathrm{qNIE}}
\newcommand{\QTE}{\mathrm{qTE}}

\usepackage{cleveref}
\crefname{theo}{Theorem}{Theorems}
\crefname{lemm}{Lemma}{Lemmas}
\crefname{coro}{Corollary}{Corollaries}
\crefname{prop}{Proposition}{Propositions}

\Crefname{condition}{Condition}{Conditions}

\newlist{conditions}{enumerate}{10}
\setlist[conditions]{label*=(C\arabic*)}
\crefname{conditionsi}{condition}{conditions}
\Crefname{conditionsi}{Condition}{Conditions}
\crefrangeformat{conditionsi}{conditions~#3#1#4--#5#2#6}
\crefmultiformat{conditionsi}%
{conditions~#2#1#3}{ and~#2#1#3}{, #2#1#3}{ and~#2#1#3}
\crefrangemultiformat{conditionsi}%
{conditions~#3#1#4--#5#2#6}{ and~#3#1#4--#5#2#6)}{, #3#1#4--#5#2#6}{ and~#3#1#4--#5#2#6}

\Crefrangeformat{conditionsi}{Conditions~#3#1#4--#5#2#6}
\Crefmultiformat{conditionsi}%
{Conditions~#2#1#3}{ and~#2#1#3}{, #2#1#3}{ and~#2#1#3}
\Crefrangemultiformat{conditionsi}%
{Conditions~#3#1#4--#5#2#6}{ and~#3#1#4--#5#2#6)}{, #3#1#4--#5#2#6}{ and~#3#1#4--#5#2#6}

\newcommand{\E}{\mathrm{E}}

\newcommand{\var}{\textrm{var}}

\newcommand{\todistribution}{\xrightarrow{\text{d}}}

\newcommand{\boottodistribution}{\stackrel{\text{d}^\ast}{\rightsquigarrow}}

\newcommand{\mZ}{\mathbb{Z}}

\newcommand{\LT}{\mathrm{LT}}

 \usepackage{adjustbox}
\usepackage{xr}
\makeatletter
\newcommand*{\addFileDependency}[1]{
	\typeout{(#1)}
	\@addtofilelist{#1}
	\IfFileExists{#1}{}{\typeout{No file #1.}}
}\makeatother

\newcommand*{\myexternaldocument}[1]{%
	\externaldocument{#1}%
	\addFileDependency{#1.tex}%
	\addFileDependency{#1.aux}%
}
\myexternaldocument{supp}

\usepackage{amsmath}

\usepackage{newtxtext}
\usepackage[subscriptcorrection]{newtxmath}

\graphicspath{{./figs/}}

\usepackage[plain,noend]{algorithm2e}

\makeatletter
\renewcommand{\algocf@captiontext}[2]{#1\algocf@typo. \AlCapFnt{}#2} 
\def\@algocf@capt@plain{top}
\renewcommand{\algocf@makecaption}[2]{%
  \addtolength{\hsize}{\algomargin}%
  \sbox\@tempboxa{\algocf@captiontext{#1}{#2}}%
  \ifdim\wd\@tempboxa >\hsize
    \hskip .5\algomargin%
    \parbox[t]{\hsize}{\algocf@captiontext{#1}{#2}}
  \else%
    \global\@minipagefalse%
    \hbox to\hsize{\box\@tempboxa}
  \fi%
  \addtolength{\hsize}{-\algomargin}%
}
\makeatother

\begin{document}

\jname{Biometrika}
\jyear{2024}
\jvol{103}
\jnum{1}
\cyear{2024}
\accessdate{Advance Access publication on 31 July 2023}

\received{2 January 2017}
\revised{1 August 2023}

\markboth{C. Chen et~al.}{Quantile Mediation Analytics}

\title{Quantile Mediation Analytics}

\author{Canyi Chen}
\affil{Department of Biostatistics, University of Michigan,\\ Ann Arbor, Michigan 48104, U.S.A.
\email{canyic@umich.edu}}

\author{Yinqiu He}
\affil{Department of Statistics, University of Wisconsin,\\ Madison, Wisconsin 53706, U.S.A. \email{yinqiu.he@wisc.edu}}

\author{Huixia J. Wang}
\affil{Department of Statistics, George Washington University,\\ Washington, DC 20052, U.S.A.
\email{judywang@email.gwu.edu}}

\author{Gongjun Xu}
\affil{Department of Statistics, University of Michigan,\\ Ann Arbor, Michigan 48104, U.S.A.
	\email{gongjun@umich.edu}}

\author{\and Peter X.-K. Song}
\affil{Department of Biostatistics, University of Michigan, \\Ann Arbor, Michigan 48104, U.S.A. \email{pxsong@umich.edu}}

\maketitle

\begin{abstract}
Mediation analytics help examine if and how an intermediate variable mediates the influence of an exposure variable on an outcome of interest. Quantiles, rather than the mean, of an outcome are scientifically relevant to the comparison among specific subgroups in practical studies. Albeit some empirical studies available in the literature, there lacks a thorough theoretical investigation of quantile-based mediation analysis, which hinders practitioners from using such methods to answer important scientific questions. To address this significant technical gap, in this paper, we develop a quantile mediation analysis methodology to facilitate the identification, estimation, and testing of quantile mediation effects under a hypothesized directed acyclic graph.  We establish two key estimands, quantile natural direct effect (qNDE) and quantile natural indirect effect (qNIE), in the counterfactual framework, both of which have closed-form expressions. To overcome the issue that the null hypothesis of no mediation effect is composite, we establish a powerful adaptive bootstrap method that is shown theoretically and numerically to achieve a proper type I error control. We illustrate the proposed quantile mediation analysis methodology through both extensive simulation experiments and a real-world dataset in that we investigate the mediation effect of lipidomic biomarkers for the influence of exposure to phthalates on early childhood obesity clinically diagnosed by 95\% percentile of body mass index.
\end{abstract}

\begin{keywords}
Adaptive bootstrap; causal estimand; composite null hypothesis; Gaussian copula; generalized structural equation model. 
\end{keywords}

\addtolength{\textheight}{.5in}%
\section{Introduction\label{section:introduction}}
Mediation analysis is one of the effective statistical methodologies to learn and infer structural parameters in a certain pathway of scientific importance. It enables practitioners to determine if, and to what extent, one or many immediate variables mediate the influence of exposures, either environmental or social, on outcomes of interest under hypothesized prespecified directed acyclic graphs (DAGs); see, for example,  \cite{sobel1982AsymptoticConfidenceIntervals} and    \cite{baron1986ModeratorMediatorVariable}. 
To date, mediation analysis has been extensively applied in practice, such as psychology, 
 genomics and epidemiology \citep{guo2022HighDimensionalMediationAnalysisa}, just name a few.

Most existing theories, methods, and applications of mediation analysis have concerned the mean of an outcome with few results available in the quantile regression paradigm. 
However, studying quantiles is crucial as they offer a unique perspective for examining the data-generation distribution, addressing important scientific questions that the mean alone may not fully capture. For example, in studies of childhood obesity, children aged two or older are clinically diagnosed as obese if their body mass index (BMI) falls at or above the 95th percentile for their age \citep{centersfordiseasecontrolandprevention2022ChildTeenBMI}. In aging studies, an individual with their biological age surpassing the population median is deemed as accelerated aging, a {stage of evolution} linked to elevated risk of diseases onset such as Alzheimer's, diabetes and cancers \citep{horvath2013DNAMethylationAge}. 

Three significant technical challenges arise when quantiles, rather than the mean, of an outcome are of interest in mediation analyses. First, the continuity of the variables disqualifies existing stratification-based methods proposed by \cite{huber2022DirectIndirectEffects} while discretization approaches \citep{hsu2023DoublyRobustEstimation} may suffer from approximation errors and/or excessive computational costs in large-scale applications. Second, in comparison with the linear mean model, the nonlinearity of the quantile regression introduces inherent complexity to establish identifiability conditions for causal effects, resulting in a paucity of literature offering interpretable expressions for quantile mediation effects \citep{shen2014QuantileMediationModels,yuan2014RobustMediationAnalysis,bind2017QuantileCausalMediation}. Third, partly arising from the first two, involves the development of valid hypothesis testing procedures. As we will demonstrate in \Cref{section:adaptive_bootstrap}, since the null hypothesis of no quantile mediation effect appears composite,  the limiting null distributions are found to vary across different subspaces of the null hypothesis. Consequently, conservative control of type I error prevails in almost all existing tests, including popular generalized Sobel's test and MaxP test as well as the classical bootstrap procedure \citep{sobel1982AsymptoticConfidenceIntervals,yuan2014RobustMediationAnalysis,wang2023NonequivalenceTwoLeastabsolutedeviation}. Such conservatism is illustrated by both simulation studies and data applications in  \Cref{section:simulation,section:real_data}.

In this paper, we propose an important extension of generalized structural equation models \citep[SEM]{hao2023ClassDirectedAcyclic} to set up a unified framework for evaluating quantile mediation effects in the presence of continuous exposure, mediator and outcome variables. Through such a new framework, we can write key quantile mediation estimands in closed-form expressions so that analyzing and interpreting quantile mediation effects becomes conceptually straightforward and computationally easy.  In addition, we introduce a valid hypothesis testing procedure based on an adaptive bootstrap (AB) to test the null hypothesis of no quantile mediation effect. Our AB test leverages the explicit derivation of the limiting null distribution under the composite null, followed by a pretest to distinguish among different null hypotheses prior to conducting the bootstrap. The consistency of the bootstrap procedure under both the null and a local alternative ensures valid Type I error control with enhanced statistical power.

Of note, the generalized SEM is originally proposed to  analyze mean mediation effects in non-Gaussian data, using hierarchical modeling and Gaussian copula to construct the joint distribution. An advantage of employing Gaussian copula in the construction of DAG stems from the motivation of preserving causal topology in a similar spirit to the classical linear SEM \citep{hao2023ClassDirectedAcyclic,wang2018CopulaBasedQuantileRegression,zhou2021StructuralFactorEquation}. Refer to a further discussion of the DAG topology preservation in \Cref{section:generalized_SEM}, where the causal pathway is exclusively characterized by the second moments of the joint distribution while confounding variables affect only the first moments of the joint distribution \citep{frisch1933PartialTimeRegressions}. In such constructed DAG, quantiles naturally hedge with rank-based correlations for easy interpretation. Furthermore, because of such separability between DAG topology and confounding adjustment, the resulting quantile mediation analytics appear to be rather resilient to complex data structures and lead to interpretable results of practical relevance. Moreover, the model diagnostics are readily available to check specific model assumptions \citep{zhang2016GoodnessoffitTestSpecification} in addition to conventional sensitivity analyses.

This paper is organized as follows.  \Cref{section:quantile_mediation_effect} introduces the generalized SEM, followed by the derivation of key quantile mediation estimands, where more details of the generalized SEM are relegated to  \Cref{section:generalized_SEM}.  \Cref{section:adaptive_bootstrap} concerns the development of adaptive bootstrap procedures for hypothesis testing of composite nulls. \Cref{section:simulation,section:real_data} focus on evaluating the performance of the proposed methodology through, respectively,  extensive simulation experiments and a real data application. Concluding remarks are included in \Cref{section:conclusion}. All technical proofs are relegated to the online Supplementary Material.


\section{Quantile Mediation Effect}\label{section:quantile_mediation_effect}

\begin{figure}[tbph!]
	\centering{\renewcommand{\arraystretch}{0.5} 
		\begin{tabular}{cc}
	\begin{tikzpicture}
	\node[draw, circle] (S) at (0, 0) {$S$};
	\node[draw, circle] (M) at (2, 1.5) {$M$};
	\node[draw, circle] (Y) at (4, 0) {$Y$};
	
	\draw[->] (S) -- (M) node[midway, below] {$\alpha_S$};
	\draw[->] (M) -- (Y) node[midway, below] {$\beta_M$};
	\draw[->] (S) -- (Y) node[midway, below] {$\gamma_S$};
\end{tikzpicture}

			&     
	\begin{tikzpicture}
	\node[draw, circle] (S) at (0, 0) {$S$};
	\node[draw, circle] (M) at (2, 1.5) {$M$};
	\node[draw, circle] (Y) at (4, 0) {$Y$};
	\node[draw, circle] (X) at (2, 3) {$\X$};
	
	\draw[->] (S) -- (M) node[midway, below] {$\alpha_S$};
	\draw[->] (M) -- (Y) node[midway, below] {$\beta_M$};
	\draw[->] (S) -- (Y) node[midway, below] {$\gamma_S$};
	\draw[->] (X) -- (S);
	\draw[->] (X) -- (M);
	\draw[->] (X) -- (Y);
\end{tikzpicture}\\
			(A)   & (B)
		\end{tabular}
	}
	\captionsetup{font = footnotesize}
	\caption{
		Directed acyclic graphs for unconfounded and confounded mediation models in (A) and (B), respectively. 
		$S$: exposure, $M$: mediator,  $Y$: outcome, $\X$: confounders; $\alpha_S, \beta_M, \gamma_S$:  structural coefficients.
	}\label{fig:mediationDAG_unconfounded}
\end{figure}
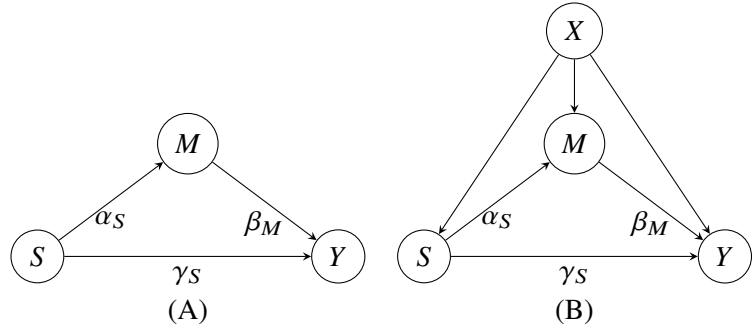

We begin with the generalized SEM, a joint distribution of exposure $S\in\mR$, mediator $M\in\mR$, outcome $Y\in\mR$ conditional on the $p$-element vector of confounders $\X\in\mR^p$ given by
\beqr\label{model:GSEM}
F_{S, M, Y\mid \X}(s, m, y\mid \x) = C\{F_{S\mid \X}(s\mid \x), F_{M\mid \X}(m\mid \x), F_{Y\mid \X}(y\mid \x); \bGam^\prime\},
\eeqr
where $C$ is a Gaussian copula \citep{song2000MultivariateDispersionModels}, and the dependence matrix $\bGam^\prime$ is given in (\ref{model:bGam}); see \Cref{section:generalized_SEM} for more details. Such specified joint distribution $F_{S, M, Y\mid \X}(s, m, y\mid \x)$ incorporates generalized linear models (GLMs) in the marginal distributions, respectively.   According to \cite{hao2023ClassDirectedAcyclic}, this construction preserves the DAG topology shown in \Cref{fig:mediationDAG_unconfounded} (B) with two key pathways, namely $S\rightarrow Y$ and $ S \rightarrow M \rightarrow Y$. In classical linear SEM, the former pathway refers to a direct effect $\gamma_S$, while the latter is known as the mediation pathway with effect size $\alpha_S \beta_M$. These effects become more complicated in the generalized SEM in connection to quantiles, which will be studied in detail in this paper.  This joint distribution can produce conditional distributions, marginal distributions, and, more importantly, quantile functions as desired for the study of quantile mediation effects. 


\subsection{Definition of Quantile Causal Effects}
Given a quantile level $\tau\in(0, 1)$ and random variables $S$ and $Y$, $Q_Y(\tau)\defn \inf\{y\in\mR\colon \pr(Y\leq y)\geq \tau\}$ denotes its $\tau$-th  marginal quantile of $Y$, and $Q_{Y\mid S}(\tau\mid s) \defn \inf\{y\in\mR\colon \pr(Y\leq y\mid S = s)\geq \tau\}$ denotes the $\tau$-th conditional quantile of $Y$ on variable $S = s$. 
In the following, $Q_{Y|S}(\tau|s)$ is abbreviated as $Q_{Y|S}(\tau)$ when there is no ambiguity. 



To proceed with the counterfactual paradigm, let $M(s)$ represent a potential value of the mediator $M$ when exposure $S=s$, and let $Y(s, m)$ represent a potential outcome when exposure $S=s$ and mediator $M=m$. We assume the Stable Unit Treatment Value assumption \citep[SUTV,][]{rubin1980RandomizationAnalysisExperimental}:
\begin{condition}[Stable Unit Treatment Value Assumption]  \label{condition:SUTV} $M = M(S),\enskip \text{and}\enskip Y = Y(S, M(S))$.
\end{condition}
Given two exposures levels $s$ and $s^\prime$, the {conditional quantile natural direct effect (qNDE)} at the $\tau$-th quantile level is defined as,
\begin{equation}
\QNDE_\tau(s, s^\prime;  \x)\defn Q_{Y(s^\prime, M(s))\mid \X}(\tau\mid \x) - Q_{Y(s, M(s))\mid \X}(\tau\mid \x).
\label{eq:qNDE}
\end{equation}
Likewise, the conditional quantile natural indirect  effect (qNIE) at the $\tau$-th quantile level is defined as
\begin{equation}
\QNIE_\tau(s, s^\prime; \x)\defn Q_{Y(s^\prime, M(s^\prime))\mid \X}(\tau\mid \x) - Q_{Y(s^\prime, M(s))\mid \X}(\tau\mid \x). 
\label{eq:qNIE}
\end{equation} 
It follows that a sum of qNDE and qNIE gives the conditional quantile total effect (qTE) at the $\tau$-th quantile level:
\beqrs
\mathrm{qTE}_{\tau}(s, s^\prime; \x) & \defn & \QNIE_\tau(s, s^\prime;  \x) +  \QNDE_\tau(s, s^\prime;  \x) =  Q_{Y(s^\prime, M(s^\prime))\mid \X}(\tau\mid \x) -  Q_{Y(s, M(s))\mid \X}(\tau\mid \x). 
\eeqrs
This $\mathrm{qTE}_\tau$ is similar to the $\eta = \alpha_S\beta_M + \gamma_S$ in the classical SEM introduced in \Cref{section:generalized_SEM}, and summarizes all causal impacts on $Y$ at the $\tau$-th quantile level due to $S$. 

The above definitions of  conditional qNDE and qNIE present a substantial extension of the conditional quantile total effect discussed in \cite{li2023EvaluatingDynamicConditional}, with the inclusion of a mediator and associated pathways. The adoption of the copula hierarchical modeling approach provides two key technical advantages. First, it significantly simplifies the estimation procedure as the DAG parameters and effects of confounding factors can be separately estimated, thereby greatly lowering computational demands. Second, it can improve estimation efficiency and statistical power in the inference for qNDE and qNIE by minimizing the disturbance from confounders. 
Moreover, both qNDE and qNIE defined under the unified framework of generalized SEM give attrative interpretability comparable to that of their counterparts in the classical linear SEM. Thus, our methodology covers many important models that have been studied separately in the literature, including  \cite{vanderweele2010OddsRatiosMediationa} examining conditional odds ratios, and \cite{bind2017QuantileCausalMediation} exploring conditional controlled effects in the context of longitudinal data. 

\subsection{Identifiability}
We now present sufficient conditions for the identification of causal quantile mediation effect $S\to M\to Y$ as shown in \Cref{fig:mediationDAG_unconfounded} under the counterfactual framework \citep{vanderweele2015ExplanationCausalInference}. These conditions are widely utilized in the causal inference literature to ensure that the causal effect can be expressed as a function of observable quantities.   One key theoretical advantage is that under the assumption of {\it Sequential Ignorability} \citep[SI,][]{imai2010IdentificationInferenceSensitivity,shpitser2011CompleteGraphicalCriterion}, the generalized SEM provides closed-form expressions for both  $\QNIE_\tau(s, s^\prime; \x)$ and $\QNDE_\tau(s, s^\prime; \x)$, which brings not only technical convenience and numerical ease but also interpretability in practice.
\begin{condition}[Sequential Ignorability Assumption]
	  \label{condition:sequential_ignorability} For all levels of $s$, $s^\prime$, $m$ and $\x$, the poential outcomes generated by the generalized SEM satisfy:  (i) $\{Y(s, m), M(s^\prime)\}\independent S \mid \X = \x$; and  
	(ii)  $Y(s, m)\independent  M(s^\prime)\mid \{S = s^\prime, \X = \x\}$.
\end{condition}
Here $A\independent B\mid C$ denotes the conditional independence of $A$ and $B$ on $C$.  
\Cref{condition:sequential_ignorability} (i) states that given confounders $\X$, exposure $S$ is independent of the potential outcome $Y(s, m)$ and the potential mediator $M(s^\prime)$.  \Cref{condition:sequential_ignorability} (ii) requires that the potential outcome $Y(s, m)$ and the mediator $M$ are independent conditional on $S$ and $\X$. Under the above identifiability conditions, both qNDE and qNIE can be written as closed-form expressions in \Cref{theorem:explicit_form_continuous_case} below.  For any vaue $s$ of $S$, denote the normal score by $z_{s} \equiv z_{s}(\x) \defn \Phi^{-1}\{F_{S\mid\X}(s\mid \x)\}$, and let $\delta_Y = (\eta^2 + \beta_M^2 + 1)^{1/2}$ with $\eta = \alpha_S\beta_M + \gamma_S$.

{\theorem  \label{theorem:explicit_form_continuous_case}
For continuous $S$, $M$ and $Y$,  under \Cref{condition:SUTV,condition:sequential_ignorability}, the closed-form expressions of $\QNIE_\tau(s, s^\prime; \x)$ and $\QNDE_\tau(s, s^\prime; \x)$ are given by, respectively,
\beqrs
\QNDE_\tau(s, s^\prime;   \x)
& = & Q_{Y\mid \X}\{\Phi(\Delta_{s^\prime, s}(\tau))\mid \x\}  - Q_{Y\mid \X}\{\Phi(\Delta_{s, s}(\tau))\mid \x\},\enskip \text{and}\\
\QNIE_\tau(s, s^\prime;   \x) 
& = & Q_{Y\mid \X}\{\Phi(\Delta_{s^\prime, s^\prime} (\tau))\mid \x\}  - Q_{Y\mid \X}\{\Phi(\Delta_{s^\prime, s} (\tau))\mid \x\},
\eeqrs
where $\Delta_{s^\prime, s} (\tau) \equiv \Delta_{s^\prime, s}(\x;\tau) \defn \{\gamma_S z_{s^\prime} + \alpha_S \beta_M z_{s} + \Phi^{-1}(\tau)(1 + \beta_M^2)^{1/2}\}/\delta_Y$ and other terms are similarly defined. 
Moreover, we have (i) $\QNDE_\tau(s, s^\prime;   \x) = 0$ if and only if $\gamma_S = 0$;  and (ii) $\QNIE_\tau(s, s^\prime;   \x) = 0$ if and only if $\alpha_S\beta_M = 0$. 
}

The term $\Delta_{s^\prime, s} (\tau)$ plays a pivotal role in both $\QNDE_\tau(s, s^\prime; \x)$ and $\QNIE_\tau(s, s^\prime; \x)$, which is comprised of two components.  One is the constant base component, $(\gamma_S z_{s^\prime} + \alpha_S \beta_M z_{s})/\delta_Y$, which is independent of quantile level $\tau$, and the other is ``drift" component depending multiplicatively on the $\tau$-th normal quantile, $\Phi^{-1}(\tau)(1 + \beta_M^2)^{1/2}/\delta_Y$, . The base component is driven by the exposure levels $s$ and $s^\prime$ along with the strengths of two pathways $(\alpha_S\beta_M, \gamma_S)$. In contrast, the drift is proportionally modified by the $\tau$-level normal quantle on $\{1 - {\eta^2}/{(\eta^2 + \beta_M^2 + 1)} \}^{1/2} = [1-\mbox{corr}^2\{z_S(\x), z_Y(\x)\}]^{1/2}$, which may be regarded as ``the unexplained effect by $S$" for outcome $Y$ with the generalized SEM approach to model the DAG.  While this decomposition shares similarities with the quantiles of the Gaussian distribution, the non-linear rank transformation $\Phi^{-1}{F_{S \mid \X}(\cdot \mid \x)}$ allows our framework to accommodate more general data types.

%


\Cref{theorem:explicit_form_continuous_case} depicts profiles of qNIE and qNDE across different quantile levels for many concrete examples. Below we  exemplify \Cref{theorem:explicit_form_continuous_case} with two particular examples.

{\example\label{corollary:reduced_to_average_mediation_effect_under_marginal_normality}
When $S$ and $Y$ given $\X = \x$ are normally distributed,  $S\mid \X = \x\sim\calN(\mu_S(\x), \sigma_S^2)$ and $Y\mid \X = \x \sim \calN(\mu_Y(\x), \sigma^2_Y)$, 
it is easy to show that the $\QNDE_\tau$ and $\QNIE_\tau$ are given by
\beqrs
\QNDE_\tau(s, s^\prime; \x) =  \frac{\sigma_Y}{\sigma_S\delta_Y}   \gamma_S (s^\prime  - s),\enskip\text{and}\enskip \QNIE_\tau(s, s^\prime; \x) =  \frac{\sigma_Y}{\sigma_S\delta_Y} \alpha_S \beta_M (s^\prime - s).
\eeqrs
This example is trivival in the sense that both estimands are independent of quantile level $\tau$ and in fact coincide with their mean counterparts given in \cite{hao2023ClassDirectedAcyclic}. The normal quantile term $\Phi^{-1} (\tau)$ is canceled out in calculation of the qNDE and qNIE.  
}

	
	Such cancellation disappears in many cases. The following example, motivated by our childhood obesity metabolic study in \Cref{section:real_data}, illustrates a nontrivial application of qNDE and qNIE.

\begin{figure}[htbp!]
\centerline{\renewcommand{\arraystretch}{0.8} 
	\begin{tabular}{cc}			
		\psfig{figure=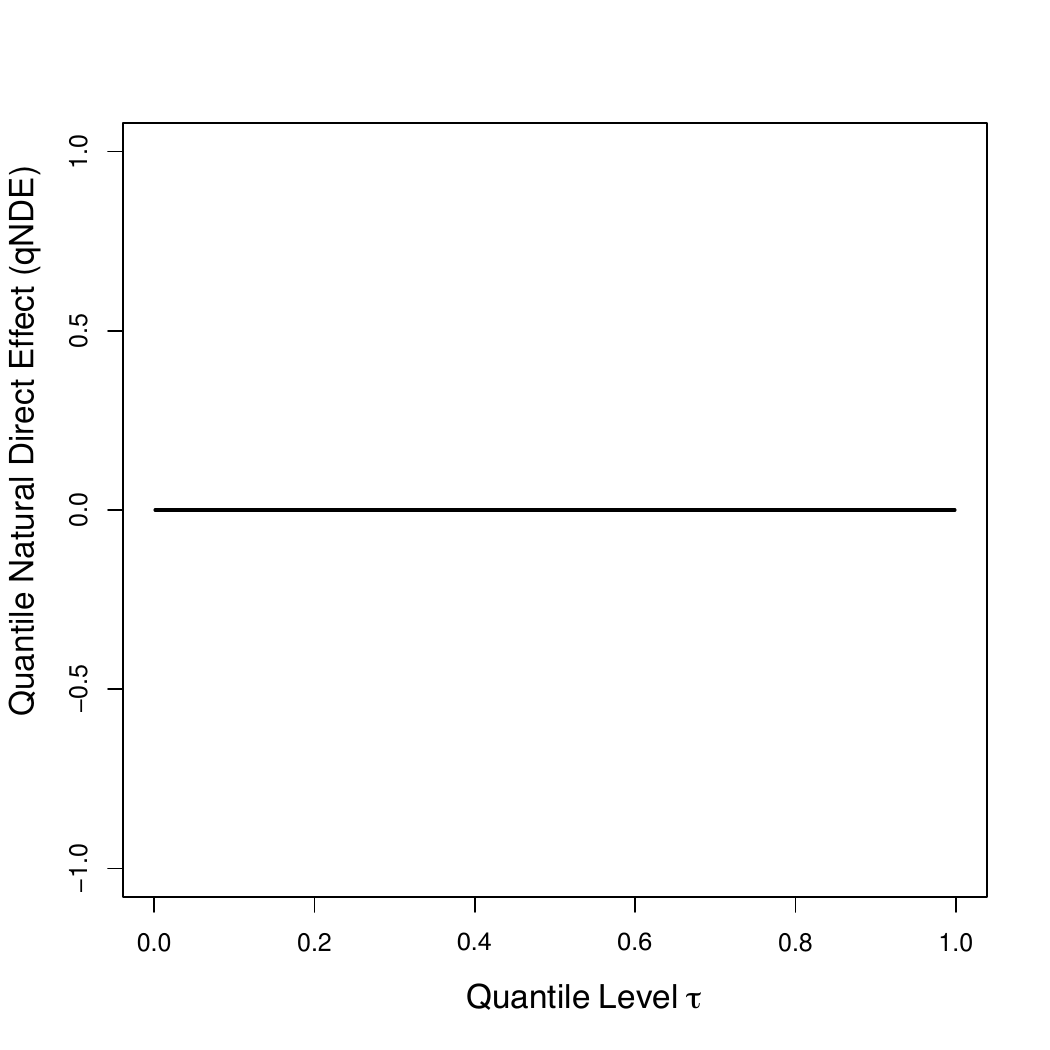,width=2.7in,angle=0} & \psfig{figure=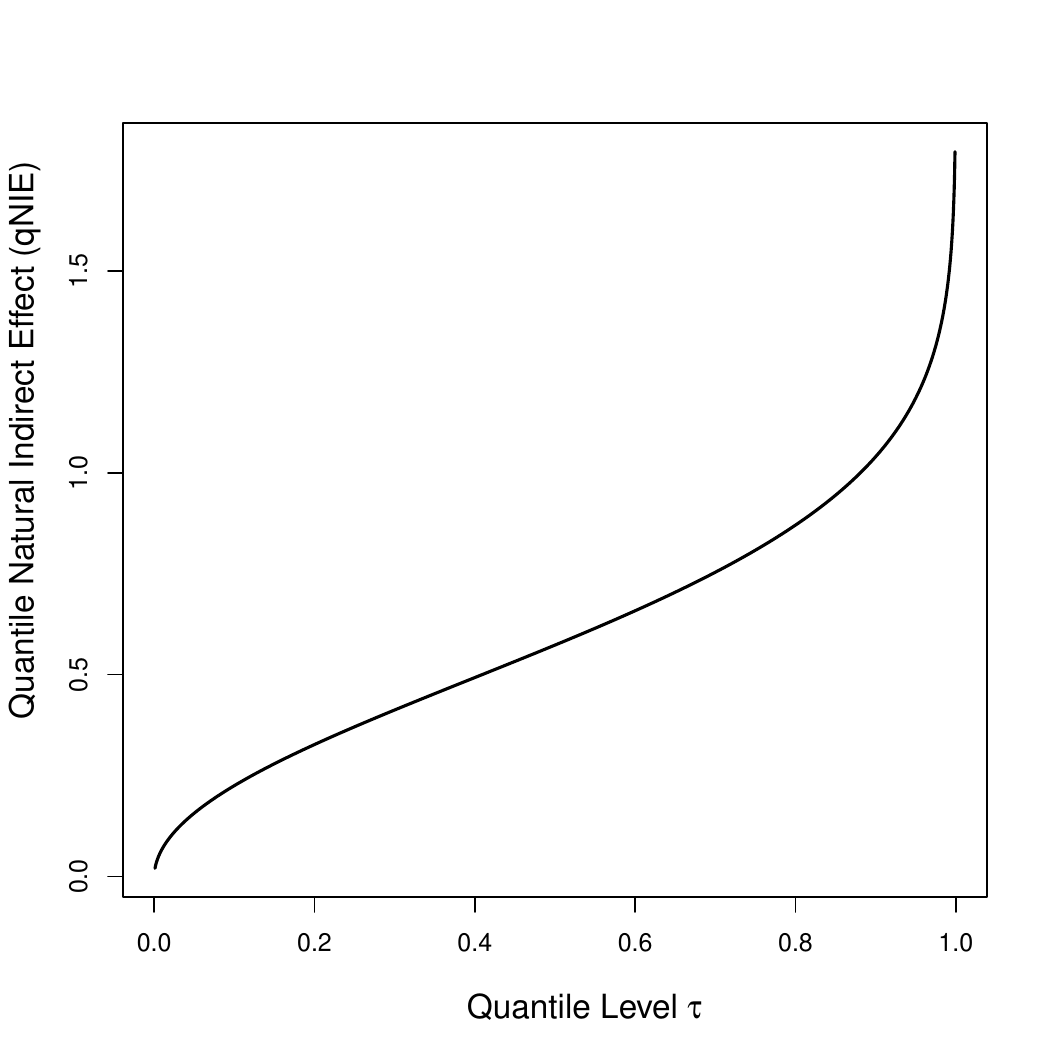,width=2.7in,angle=0}\\
		(A): Quantile natural direct effect   & (B): Quantile natural indirect  effect
	\end{tabular}
}
\captionsetup{font = footnotesize}
\caption{ Plots of the qNDE in (A) and qNIE in (B) under a generalized SEM with $X\sim\calN(0, 1)$, $M\sim\calN(0, 1)$ and $Y\sim\text{Exp}(1)$. The DAG model parameters $(\alpha_S, \beta_M, \gamma_S)$ are set so that qNDE equals to zero (constant) while qNIE varies over quantile level $\tau$ when exposure changes from $s = 0$ to $s^\prime = 1$. }
\label{fig:qNDE_qNIE_normal_exp}
\end{figure}

	\begin{example}\label{example:heterogeneity}
For simplicity, we consider a special scenario where the $\QNDE_\tau(s, s^\prime; \x)$ is zero, but the $\QNIE_\tau(s, s^\prime; \x)$ varies over $\tau$, and $\X \equiv 1$ (no confouding). Suppose that marginally $S\sim\calN(0,1)$, $M\sim\calN(0,1)$ and $Y\sim\mathrm{Exp}(1)$, and the joint distribution of $(S, M, Y)$ is given by the generalized SEM via the Gaussian copula. Set $(\alpha_S, \beta_M, \gamma_S) =(1, 1, 0)$. Consider a change from $s=0$ to $s^\prime =1$. We plot $\QNDE_\tau(s, s^\prime; \x)$ and  $\QNIE_\tau(s, s^\prime; \x)$ over $\tau$, respectively, in \Cref{fig:qNDE_qNIE_normal_exp} (A) and (B). The plot of qNDE is flat because we intensionally set $\gamma_S = 0$,  and the plot of $\QNIE_\tau(s, s^\prime; \x)$ is an increasing function in $\tau$ under $\alpha_S\beta_M\neq 0$, confirming the if-and-only-if result in \Cref{theorem:explicit_form_continuous_case}.
\end{example}

\subsection{Analytic Tasks}
One primary task pertains to parameter estimation. Given that the generalized SEM is fully parametric, the maximum likelihood estimation (MLE) is a natural choice of the method for parameter estimation. With the availability of likelihood, we can directly estimate $\QNIE_\tau(s, s^\prime; \x)$ and $\QNDE_\tau(s, s^\prime; \x)$ by plugging in the MLEs of the model parameters. This is because both $\wh Q_{Y\mid \X}(\cdot)$ and $\wh F_{S\mid \X}(\cdot)$ are obtained as the plug-in estimates for $Q_{Y\mid \X}(\cdot)$ and $F_{S\mid \X}(\cdot)$, respectively. See the details in \Cref{appendix:detailed_form} of the Supplementary Material.

In the implementation of MLE, we adopt the widely used strategy proposed by \cite{joe2005AsymptoticEfficiencyTwostage} in the literature of copula dependence models. The so-called Inference Function with Marginals (IFM) is known to be computationally efficient to handle nuisance parameters in the marginals.  Specifically, given a random sample $\{(S_i, M_i, Y_i, \X_i)\}_{i = 1}^n$ of size $n$, we run MLE to obtain the estimates of the model parameters $(\wh\alpha_S, \wh\beta_M, \wh\gamma_S, \wh\bbeta_S\trans, \wh\bbeta_M\trans, \wh\bbeta_Y\trans, \wh\phi_S, \wh\phi_M, \wh\phi_Y)$ where the DAG parameter estimates  $(\wh\alpha_S, \wh\beta_M, \wh\gamma_S)$ are calculated after the marginal model parameter estimates $(\wh\bbeta_S\trans, \wh\bbeta_M\trans, \wh\bbeta_Y\trans, \wh\phi_S, \wh\phi_M, \wh\phi_Y)$ are obtained. As shown by \cite{joe2005AsymptoticEfficiencyTwostage}, IFM produces asymptotically efficient MLE.

Another primary task in the mediation analysis is to establish mediation pathways. To do so, for given $\tau$,   $s$, $s^\prime$ and $\x$,  we plan to perform a hypothesis testing of the form: 
\beqr\label{hypothesis}
H_0\colon\QNIE_\tau(s, s^\prime; \x) = 0\enskip \text{versus}\enskip H_1\colon\QNIE_\tau(s, s^\prime; \x)\neq 0.
\eeqr
Here, we focus on a single quantile level at each test, which is the most basic test considered in practice. In our motivating example with the outcome of child's BMI, we are interested in the quantile level $\tau = 0.95$ that corresponds to the clinical definition of childhood obesity. 

According to \Cref{theorem:explicit_form_continuous_case}, the null hypothesis $H_0\colon\QNIE_\tau(s, s^\prime; \x) = 0$ is equivalent to $\alpha_S\beta_M = 0$. Under this null hypothesis, the DAG parameters ($\alpha_S, \beta_M$) control all quantile profiles. In contrast, under the alternative hypothesis, the $\QNIE_\tau$ can vary with different quantile levels $\tau$. However, this hypothesis test is nontrivial due to a composite null hypothesis. That is, 
the underlying null parameter space $\Omega_0 = \{(\alpha_S, \beta_M)\colon \alpha_S\beta_M = 0\}$ constitutes three subspaces:
\beqrs
&  \Omega_{0,1} = \{(\alpha_S, \beta_M)\colon \alpha_S = 0, \beta_M \neq 0\},\enskip  \Omega_{0,2} = \{(\alpha_S, \beta_M)\colon \alpha_S \neq 0, \beta_M = 0\},\\
&  \Omega_{0,3} = \{(\alpha_S, \beta_M)\colon \alpha_S = 0, \beta_M = 0\}. 
\eeqrs

Such a composite null, especially the singleton $\Omega_{0,3}$, poses significant technical difficulties in the development of valid testing procedures. The existing popular approaches to testing the null \eqref{hypothesis} include Wald-type tests and classic nonparametric bootstrap methods. However, neither methods can control type I error properly; see our extensive numerical studies in \Cref{section:simulation}. Under singleton $\Omega_{0, 3}$, limiting behaviors of existing statistics become analytically irregular. \Cref{prop:non_regularity} provides a rigorous theoretical justification for such irregularity. \Cref{prop:non_regularity} implies that the usual first-order delta method fails to {deliver} the legitimate limiting null distribution for \eqref{hypothesis} under $\Omega_{0, 3}$, {whereas under $\Omega_{0, 1}$ and $\Omega_{0,2}$, such delta method works to establish a valid Wald-type test for $\QNIE_\tau(s, s^\prime; \x)$.}  This is because the limiting distribution of $\wh \QNIE_\tau(s, s^\prime; \x)$ varies under different null subspaces, resulting in an irregularity issue.  
{\proposition\label{prop:non_regularity}
Under the conditions of \Cref{theorem:explicit_form_continuous_case}, the irregularity issue exists under the null hypothesis $H_0$, that is
\begin{enumerate}[(i)]       
	\item $\nabla_{\alpha_S}\QNIE_\tau(s, s^\prime; \x)\Big\vert_{\beta_M = 0} = \nabla_{\beta_M}\QNIE_\tau(s, s^\prime; \x)\Big\vert_{\alpha_S = 0} = 0$,
	\item $\nabla_{\alpha_S}\QNIE_\tau(s, s^\prime; \x)\Big\vert_{\alpha_S = 0,\beta_M \neq 0}\neq 0$, (iii) $\nabla_{\beta_M}  \QNIE_\tau(s, s^\prime; \x)\Big\vert_{\alpha_S \neq 0,\beta_M = 0}\neq 0$. 
\end{enumerate} 
}
This issue has been noticed on other occasions, such as three-way contingency table analysis and factor analysis; see, for example, \cite{glonek1993BehaviourWaldStatistics}, \cite{dufour2013WaldTestsWhen}, and \cite{drton2016WaldTestsSingular}. In particular, \cite{wang2023NonequivalenceTwoLeastabsolutedeviation} pointed out that under the linear quantile mediation analysis, the composite null structure further worsens the type I error control given that the underlying true situation is unknown.

\section{Adaptive Bootstrap Test}\label{section:adaptive_bootstrap}
The recent solution based on an adaptive bootstrap (AB) test by \cite{he2023AdaptiveBootstrapTests} in the linear SEM shed light to handle the singleton $\Omega_{0,3}$. In this section, we extended the AB test to the case of quantile mediation analysis to test the null hypothesis \eqref{hypothesis} with a proper type I error control. We begin with some notations. Given two sequences of real numbers $\{a_n\}$ and $\{b_n\}$, write  $a_n = o(b_n)$ if $a_n/b_n\to 0$. Convergence in distribution is denoted by $\todistribution$, while  $\boottodistribution$ denotes bootstrap consistency with respect to the Kolmogorov-Smirnov distance  \citep[Section 23.2,][]{vaart1998AsymptoticStatistics}. More details about these convergence modes are listed in \Cref{appendix:notations} of the Supplementary Material.

\subsection{Formulation}
The key idea of the AB approach is rooted in a local asymptotic analysis framework to address the irregularity round $(\alpha_S, \beta_M) = (0, 0)$ and to examine the asymptotic behaviors of  $\wh \QNIE_\tau(s, s^\prime; \x)$ under local alternatives. That is, we explore limiting distributions with parameters deviated in a tiny amount from the $(0,0)$. Technically, given the targeted parameters $(\alpha_S, \beta_M)$, we construct local parameters $\alpha_{S, n}\defn \alpha_S + n^{-1/2} b_{\alpha_S}$ and $\beta_{M, n}\defn \beta_M + n^{-1/2} b_{\beta_M}$, where $b_{\alpha_S}>0$ and $b_{\beta_M}>0$ are given constants. Consequently, a  generalized SEM may be formed in light of the following local covariance structure:
\beqr\label{model:local_generalized_SEM}
\bGam_n = (\I - \bTh_n)^{-1}(\I - \bTh_n)^{-{\mbox{\tiny{T}}}},\enskip \text{ with }\enskip 
\bTh_n = \LT(\alpha_{S, n}, \gamma_S, \beta_{M, n}).
\eeqr
The utility of $n^{-1/2}$-vicinity of local neighboring values $(\alpha_{S, n}, \beta_{M, n})$ is technical by appealing to proceed theoretical investigation of local asymptotic behaviors in order to test the null hypothesis \eqref{hypothesis}. This type technique has also been used by  \cite{wang2018TestingMarginalLineara} and  \cite{he2023AdaptiveBootstrapTests}. 

To proceed, we assume some regularity conditions that are widely used in the literature of M estimation and bootstrap to ensure both consistency and asymptotic normality \citep{vaart1998AsymptoticStatistics}. 

\begin{condition}
 \label{condition:mle_and_boot} Let $\btheta_0\defn (\alpha_S, \beta_M, \gamma_S, \bbeta_S\trans, \bbeta_M\trans, \bbeta_Y\trans, \phi_S, \phi_M, \phi_Y)\trans\in\mR^{3p +6}$ be the true parameters. Denote the log likelihood by $\ell_n(\btheta)\defn 1/n\sum_{i = 1}^n \ell_{\btheta}(S_i, M_i, Y_i, \X_i)$. 
	Assume that the parameter space $\bTh_0\subset\mR^{3p +6}$ that $\btheta_0$ resides is open and compact. Let $\epsilon>0$. (i) The log-likelihood function $\ell_{\btheta}$ is twice continuously differentiable almost surely and dominated by certain integrable function. (ii) In a neighbor of $\btheta_0$, $\ell_{\btheta}$ is Lipschitz continuous regarding $\btheta$ with  a Lipschitz constant   $L(s, m, y, \x)$ that has a finite $(2+\epsilon)$th moment. (iii) The score function $\nabla_{\btheta}\ell_{\btheta}$ is dominated by a  function whose   $(2+\epsilon)$th moment is finite.  (iv) The variability matrix $\E(\nabla_{\btheta}^2\ell_{\btheta})$ exists and is nonsingular at point $\btheta_0$.   
\end{condition}

{\theorem
\label{theorem:asymptotic_property}
Assume \Cref{condition:SUTV,condition:sequential_ignorability,condition:mle_and_boot} hold.   {Consider the null hypothesis $H_0$ in the generalized SEM in  \eqref{model:local_generalized_SEM}.} 
We have 
\begin{enumerate}
	\item[(i)] when $(\alpha_S, \beta_M)\neq(0, 0)$,  
	$$
	n^{1/2} \{\wh\QNIE_\tau(s, s^\prime;  \x) - \QNIE_\tau(s, s^\prime;   \x)\}\todistribution  \eta_\tau(\x)(\alpha_S Z_2 + Z_1 \beta_M)(z_{s^\prime} - z_{s})/\delta_Y;
	$$
	\item[(ii)] when $(\alpha_S, \beta_M) = (0, 0)$,  
	$$
	n \{\wh\QNIE_\tau(s, s^\prime;   \x) - \QNIE_\tau(s, s^\prime;   \x)\} \todistribution  \eta_\tau(\x)(Z_1 Z_2  +  Z_1 b_\beta + b_\alpha Z_2)(z_{s^\prime} - z_{s})/\delta_Y,
	$$
\end{enumerate}
where $Z_1$ and $Z_2$ are random variables with the same limiting distributions of $n^{1/2}(\wh\alpha_S - \alpha_S)$ and $n^{1/2}(\wh\beta_M - \beta_M)$, respectively, and
\beqrs
\eta_\tau(\x) = \frac{\phi\{q_\tau(\x)\}}{f_{Y\mid \X}(Q_{Y\mid \X}[\Phi\{q_\tau(\x)\}])},\enskip
q_\tau(\x) = \frac{\{\gamma_S z_{s^\prime}  + \Phi^{-1}(\tau)(1 + \beta_M^2)^{1/2}\}}{\delta_Y},
\eeqrs
$z_{s} = z_{s}(\x) = \Phi^{-1}\{F_{S\mid \X}(s\mid \x)\}$, $\delta_Y = (\eta^2 + \beta_M^2 + 1)^{1/2}$, and $\phi(\cdot)$ is the probability density function of standard normal distribution.

}

\Cref{theorem:asymptotic_property} suggests that the difference  $\wh\QNIE_\tau(s, s^\prime; \x) - \QNIE_\tau(s, s^\prime; \x)$ has a non-uniform limiting distribution with regards to $(\alpha_S, \beta_M)$. This non-uniformity occurs for $(\alpha_S, \beta_M) = (0, 0)$ {or $\neq 0$; they have different convergence rates, which will be used in \Cref{subsection:ab_test} to isolate $\Omega_{0, 3}$ via a pretest procedure. } It is interesting to note that in the vicinity of $(\alpha_S, \beta_M) = (0, 0)$, the limiting distribution of $n \{\wh\QNIE_\tau(s, s^\prime; \x) - \QNIE_\tau(s, s^\prime; \x)\}$ is continuously scaled by $(b_\alpha, b_\beta)$. Such local limit allows us to increase the accuracy of the finite-sample behaviors compared to the classical nonparametric bootstrap method, that ignores the local asymptotic patterns.

\subsection{Adaptive bootstrap test}\label{subsection:ab_test}
{Separating the ``troublemaker'' $\Omega_{0,3}$ from the other two subspaces $\Omega_{0, 1}$ and $\Omega_{0, 2}$ is feasible due to the different convergence rates in \Cref{theorem:asymptotic_property}, which } can be done by 
comparing the absolute values of $ \wh T_{\alpha_S} = n^{1/2}\wh\alpha_{S}/\wh\sigma_{\alpha_S}$ and $\wh T_{\beta_M} = n^{1/2}\wh\beta_{M, n}/\wh\sigma_{\beta_M}$ to certain thresholds. {Here, $\wh\sigma_{\alpha_S}$ and $\wh\sigma_{\beta_M}$ represent any consistent estimates for the population variances $\sigma_{\alpha_S}$ and $\sigma_{\beta_M}$, respectively. 
We propose the following ``mixture'' quantity:  
\beqr\label{equation:decomposition}
\nonumber& \wh\QNIE_\tau(s, s^\prime; \x) - \QNIE_\tau(s, s^\prime; \x) = \{\wh\QNIE_\tau(s, s^\prime; \x) - \QNIE_\tau(s, s^\prime; \x) \} \times (1 - I_{\alpha_S, \lambda_n}I_{\beta_M, \lambda_n}) \\
&\hspace*{5.3cm} + \{\wh\QNIE_\tau(s, s^\prime; \x) - \QNIE_\tau(s, s^\prime; \x)\} \times I_{\alpha_S, \lambda_n}I_{\beta_M, \lambda_n},
\eeqr
where $I_{\alpha_S, \lambda_n} = I(\abs{\wh T_{\alpha_S, n}}\leq \lambda_n, \alpha_S = 0)$ and $I_{\beta_M, \lambda_n} = I(\abs{\wh T_{\beta_M, n}}\leq \lambda_n, \beta_M = 0)$ are two flags for the null subspcases, and $I(\cdot)$ is the usual indicator function. Clearly, when $(\alpha_S, \beta_M)\neq (0, 0)$, the classical bootstrap is consistent for the first term in \eqref{equation:decomposition}, while for the second term, \Cref{theorem:asymptotic_property} navigates us to construct a consistent bootstrap procedure.

As usual, a superscript $\ast$ indicates nonparametric bootstrap. When $(\alpha_S, \beta_M)= (0, 0)$, following the limit  \Cref{theorem:asymptotic_property} (ii), we may construct a bootstrap statistic $\mR^\ast_n(b_\alpha, b_\beta)$ as a bootstrap counterpart of $\eta_\tau(\x)(Z_1 Z_2  +  Z_1 b_\beta + b_\alpha Z_2)(z_{s^\prime} - z_{s})/\delta_Y$. That is,
\beqrs
\mR^\ast_n(b_\alpha, b_\beta) = \{ \mZ_1^\ast \mZ_2^\ast(\wh z_{s^\prime}^\ast -\wh z_{s}^\ast) + \mZ_1^\ast b_\beta(\wh z_{s^\prime}^\ast - \wh z_{s}^\ast) + b_\alpha \mZ_2^\ast(\wh z_{s^\prime}^\ast -\wh z_{s}^\ast)\}\wh\eta_\tau^\ast(\x)/\wh\delta_Y^\ast,
\eeqrs
where {$\wh z_{s}^\ast = \Phi^{-1}\{\wh F_{S\mid \X}^\ast(s\mid  \x)\}$}, $\wh \delta_Y^\ast = \{(\wh\gamma_S^\ast + \wh\alpha_S^\ast\wh\beta_M^\ast)^2 + (\wh\beta_M^\ast)^2 + 1\}^{1/2}$, $\wh\eta_\tau^\ast(\x) = {\phi\{\wh q^\ast(\x)\}}/{\wh f_{Y\mid \X}^\ast(\wh Q_{Y\mid \X}^\ast[\Phi\{\wh q^\ast(\x)\}])}$ with $\wh q^\ast(\x) = \{\gamma_S^\ast \wh z_{s^\prime}^\ast  + \Phi^{-1}(\tau)\}/\wh \delta_Y^\ast$, and $\mZ_1^\ast$ and $\mZ_2^\ast$ are the usual nonparametric bootstrap counterparts of $Z_1$ and $Z_2$. In our case, since $Z_1$ and $Z_2$ are two limiting variables of $ n^{1/2}\wh\alpha_S$ and $n^{1/2}\wh\beta_M$, we {propose} $\mZ_1^\ast$ and $\mZ_2^\ast$ as $n^{1/2}\wh\alpha_S^\ast$ and $n^{1/2}\wh\beta_M^\ast$. When $(\alpha_S, \beta_M)\neq(0, 0)$, we remain using the classical nonparametric bootstrap estimate $\wh\QNIE^\ast_\tau(s, s^\prime;   \x)$.

In our AB test, flags $I_{\alpha_S, \lambda_n}$ and $I_{\beta_M, \lambda_n}$ are replaced by their bootstrap counterparts of the forms:
\beqrs
I_{\alpha_S, \lambda_n}^\ast = I(\abs{\wh T_{\alpha_S}^\ast}\leq\lambda_n, \abs{\wh T_{\alpha_S}}\leq\lambda_n), \enskip I_{\beta_M, \lambda_n}^\ast = I(\abs{\wh T_{\beta_M}^\ast}\leq\lambda_n, \abs{\wh T_{\beta_M}}\leq\lambda_n),
\eeqrs
where {$\wh T_{\alpha_S}^\ast = n^{1/2}\wh\alpha_{S}^\ast/\wh\sigma_{\alpha_S}^\ast$ and $\wh T_{\beta_M}^\ast = n^{1/2}\wh\beta_{M}^\ast/\wh\sigma_{\beta_M}^\ast$} denote the bootstrap versions of  $\wh T_{\alpha_S} = n^{1/2}\wh\alpha_{S}/\wh\sigma_{\alpha_S}$ and $\wh T_{\beta_M} = n^{1/2}\wh\beta_{M}/\wh\sigma_{\beta_M}$. {Here, we use sample standard deviation estimates  based on $\{n^{1/2}\wh\alpha_{S}^{\ast}\}$ and $\{n^{1/2}\wh\beta_{M}^{\ast}\}$ for $\wh\sigma_{\alpha_S}^{\ast}$ and $\wh\sigma_{\beta_M}^{\ast}$, respectively.} 
Thus, following \eqref{equation:decomposition}, we propose an AB test statistic,
\beqrs
U^\ast_\tau  =  \{\wh\QNIE^\ast_\tau(s, s^\prime;   \x) - \wh \QNIE_\tau(s, s^\prime;   \x)\}\times (1 - I^\ast_{\alpha_S, \lambda_n}I^\ast_{\beta_M, \lambda_n})  + n^{-1} \mR^\ast_n(b_\alpha, b_\beta) \times I^\ast_{\alpha_S, \lambda_n}I^\ast_{\beta_M, \lambda_n}.
\eeqrs
We establish the bootstrap consistency of $U^\ast_\tau$.

{\theorem \label{theorem:consistency_adaptive_bootstrap}
Assume \Cref{condition:SUTV,condition:sequential_ignorability,condition:mle_and_boot}. Under the generalized SEM   \eqref{model:local_generalized_SEM}, if the tuning parameter $\lambda_n$ satisfies $\lambda_n = o(n^{1/2})$ and $\lambda_n\to\infty$ as $n\to\infty$, 
$
c_n U^\ast_\tau\boottodistribution c_n \{\wh\QNIE_\tau(s, s^\prime;  \x) - \QNIE_\tau(s, s^\prime;  \x)\},
$
where $c_n$ is a non-random scaling factor satisfying $c_n = n^{1/2}\cdot I\{(\alpha_S, \beta_M)\neq (0, 0)\} + n\cdot I\{(\alpha_S, \beta_M)= (0, 0)\}$.
}

\Cref{theorem:consistency_adaptive_bootstrap} indicates that the AB statistic $U^\ast_\tau$ is a consistent bootstrap estimate for the discrepancy $\wh\QNIE_\tau(s, s^\prime;  \x) - \QNIE_\tau(s, s^\prime;  \x)$ under the null model with $(b_\alpha, b_\beta) = (0, 0)$, if appropriately scaled. Additionally, for a fixed target parameter $(\alpha_S, \beta_M)$, in their neighborhoods, i.e., $(b_\alpha, b_\beta) \neq (0, 0)$, the bootstrap consistency remains valid as a smooth function of $(b_\alpha, b_\beta)$. This suggests that even a small change in the target parameters does not affect the bootstrap consistency in \Cref{theorem:consistency_adaptive_bootstrap}, and the AB test statistic $U^\ast_\tau$ behaves desirably under the local alternatives to gain statistical power.

In practice, since  $\alpha_S$ and $\beta_M$ are unknown, we do not use the scaling factor $c_n$  but directly use $U^\ast_\tau$ as the bootstrap statistic for $\wh\QNIE_\tau(s, s^\prime;  \x) - \QNIE_\tau(s, s^\prime;  \x)$. The rationale is that using $n^{1/2}U^\ast_\tau$ to bootstrap $n^{1/2}\{\wh\QNIE_\tau(s, s^\prime;  \x) - \QNIE_\tau(s, s^\prime;  \x)\}$ under $\Omega_{0,1}$ or $\Omega_{0,2}$ is equivalent to using $n U^\ast_\tau$ to bootstrap $n\{\wh\QNIE_\tau(s, s^\prime;  \x) - \QNIE_\tau(s, s^\prime;  \x)\}$ under $\Omega_{0,3}$. Therefore, the distribution of $U^\ast_\tau$ will approximate that of $\wh\QNIE_\tau(s, s^\prime;  \x) - \QNIE_\tau(s, s^\prime;  \x)$ regardless of the underlying true null case, as desired.

{\subsection{Implementation of  AB test}}
Given a nominal level denoted by $\omega$, we calculate the upper and lower $\omega/2$ quantiles of the bootstrap statistics $U^\ast_\tau$, denoted by $q_{1 - \omega/2}$ and $q_{\omega/2}$, respectively. If estimate  $\wh\QNIE_\tau(s, s^\prime; \x)$ falls outside the interval $(q_{\omega/2}, q_{1 - \omega/2})$, we reject the null hypothesis and conclude that the quantile mediation effect is statistically significant at the level $\omega$. It is worth reiterating that the primary goal is to test $H_0$ with the true underlying $\alpha_S$ and $\beta_M$. The strategy of creating $n^{1/2}$-local parameters $(\alpha_{S, n}, \beta_{M, n})$ is to investigate local asymptotic behaviors. Therefore, to test the null in \eqref{hypothesis} under the null model, we only need to calculate $U^\ast_\tau$ with $b_{\alpha_S} = b_{\beta_M} = 0$.



\subsection{Choice of the tunning parameter $\lambda_n$}
To apply  \Cref{theorem:consistency_adaptive_bootstrap}, we need to ensure $\lambda_n = o(n^{1/2})$ and increases to infinity as $n$ approaches infinity. Under this condition, the proposed thresholding method {provides} a consistent pretest for $\alpha_S = 0$ and $\beta_M = 0$ with an asymptotically negligible Type I error rate, i.e.,  $\pr(\abs{\wh T_{\alpha_S}}>\lambda_n, \abs{\wh T_{\beta_M}}>\lambda_n\mid \alpha_S = \beta_M = 0)\to 0$. In contrast, if $\lambda_n$ is bounded, the proposed AB test reduces to the traditional bootstrap method, which is known to suffer from inflated Type I error rates. In all our experiments, we chose $\lambda_n = \lambda n^{1/2}/\log n$ with $\lambda$ equal to 2, which worked strikingly well in all simulation studies. 

\subsection{Model diagonosis for copula spcification}

The generalized SEM is specified with the utility of a Gaussian copula for the DAG, under which we establish estimation and hypothesis testing methods. However, the choice of Gaussian copula may be subject to misspecification of the model. To confirm this specification of the model, we propose using a goodness of fit (GoF) test developed by \cite{zhang2016GoodnessoffitTestSpecification}. The central idea of the GoF test involves comparing the ``in-sample'' and ``out-of-sample'' pseudo-likelihoods. Applying this GoF test in our empirical study of the analysis of metabolic data on childhood obesity, as detailed in \Cref{fig:real_data:sensitivity:model_checking} of \Cref{section:real_data}, we showed that there exists little evidence in the data against the use of Gaussian copula in the specification of the generalized SEM.  In effect, the modeling of DAG on the rank-based dependence is rather powerful to capture nonlinear dependencies in our data and beyond; see, e.g., \cite{zhang2022SlicedIndependenceTest}.

\section{Simulation Studies}\label{section:simulation}
\subsection{Setup}
In this section, we conduct experiments to examine the finite-sample performance of our proposed method. The data are generated from the generalized SEM \eqref{model:GSEM}. Marginally, $S\mid \X\sim\calN(\X\trans\bbeta_S, \sigma^2_S)$, $M\mid \X\sim\calN(\X\trans\bbeta_M, \sigma^2_M)$, and $Y\mid \X\sim\mathrm{Exp}\{\exp(\X\trans\bbeta_Y)\}$. Set $\bbeta_S = (0.5, 0.2, 0.2, 0)\trans$, $\bbeta_M = (0.8, 0.3, 0.3, 0.4)\trans$, and $\bbeta_Y = (-0.2, 0.4, -0.2, 0.7)\trans$, and $\sigma_S = \sigma_M  = 0.3$. Write $\X = (1, X_2, X_3, X_4)\trans$ and $(X_2, X_3, X_4)\trans\sim\calN\{\bm{0}, 0.3^2\text{CS}(0.2)\}$, where $\text{CS}(0.2)$ is a compound symmetry matrix with correlation $0.2$.
Throughout, we vary sample size $n \in \{300, 500\}$, set the bootstrap sample size at $500$,  and set always $\gamma_S = 0.5$.  We report simulation results under $n = 300$ in the main text; the additional simulation results under $n = 500$ are similar and included in \Cref{section:additional_numerical_results} of the Supplementary Material. We examine the qNIE with $s = 0$, $s^\prime = 1$ and $\X = (1, 0, 0, 0)\trans$. We focus on the median quantile level $\tau = 0.5$ in all simulations. 

We compare our method (QMA-AB) with five competitors: (a) the naive nonparametric bootstrap for the quantile mediation effect test (QMA-B); (b) the classical nonparametric bootstrap for the PoC test \citep[PoC-B]{bind2017QuantileCausalMediation}; (c) a generalized Sobel's test \citep[PoC-YM]{yuan2014RobustMediationAnalysis}; (d) the classical nonparametric bootstrap for the joint significance test \citep[JS-B]{yuan2014RobustMediationAnalysis}; (e) a generalized joint significance test \citep[JS-YM]{yuan2014RobustMediationAnalysis}.


\subsection{Effect of sample size and structural parameters on estimation}\label{sectioin:additional_numerical_results:estimation}
\captionsetup{font=footnotesize}
In this subsection, we investigate the impact of sample size and structural parameters on estimating $\QNIE_\tau$ and $\QNDE_\tau$. The sample size $n$ varies over $\{200, 400, 600, 800, 1000\}$, while the structural parameters $(\alpha_S, \beta_M)$ are selected from different combinations in $\{0, 0.5, 0.5\}$. The mean squared error (MSE), along with the ratio of MSE relative to the MSE for $n = 200$, is summarized in \Cref{table:estimation}. The simulation results reveal distinct convergence rates, consistent with our theoretical findings in \Cref{theorem:asymptotic_property}. Notably, the MSE for $\QNIE_\tau$ decreases quadratically as the sample size increases when $(\alpha_S, \beta_M) = (0, 0)$, and decreases linearly for other parameter combinations.

\begin{table}[htpb!]
\caption{Mean squared error (MSE) of $\QNIE_\tau$ and $\QNDE_\tau$ across different combinations of $(\alpha_S, \beta_M)$ for varying sample sizes $n \in \{200, 400, 600, 800, 1000\}$. The table also presents the ratio of MSE relative to the MSE at $n = 200$. The quantile level is fixed at $\tau = 0.5$, and $\gamma_S = 0.5$.}\label{table:estimation}
\centering{\footnotesize
	\begin{tabular}{l c ccccc | cccc}
		$(\alpha_S, \beta_M)$ & $n$ & 200 & 400 & 600 & 800 & 1000 & 400 & 600 & 800 & 1000 \\ 
		& & \multicolumn{5}{c}{MSE} & \multicolumn{4}{c}{Ratio of MSE}\\
		\multirow{2}{*}{(0, 0)} & $\QNIE_\tau$ & 0.26 & 0.07 & 0.03 & 0.02 & 0.01 & 3.59 & 8.61 & 14.74 & 22.56 \\ 
		& $\QNDE_\tau$ & 21.15 & 10.48 & 6.76 & 5.11 & 4.33 & 2.02 & 3.13 & 4.14 & 4.88 \\ 
		\multirow{2}{*}{(0.5, 0)} & $\QNIE_\tau$ & 14.77 & 7.05 & 4.55 & 3.44 & 2.75 & 2.10 & 3.25 & 4.29 & 5.36 \\ 
		& $\QNDE_\tau$ & 34.41 & 16.94 & 11.43 & 8.49 & 7.00 & 2.03 & 3.01 & 4.06 & 4.91 \\ 
		\multirow{2}{*}{(0, 0.5)} & $\QNIE_\tau$ & 11.46 & 6.13 & 3.73 & 2.88 & 2.34 & 1.87 & 3.07 & 3.98 & 4.91 \\ 
		& $\QNDE_\tau$ & 15.56 & 7.90 & 5.43 & 3.89 & 3.17 & 1.97 & 2.87 & 3.99 & 4.91 \\ 
		\multirow{2}{*}{(0.5, 0.5)} & $\QNIE_\tau$ & 11.46 & 6.13 & 3.73 & 2.88 & 2.34 & 1.87 & 3.07 & 3.98 & 4.91 \\ 
		& $\QNDE_\tau$ & 15.56 & 7.90 & 5.43 & 3.89 & 3.17 & 1.97 & 2.87 & 3.99 & 4.91 \\ 
\end{tabular}}
\end{table}

\subsection{Type I error rate}\label{section:simulation:null}

{\it Setting I: A fixed null.} In the setting of fixed null, we examine the distributions of the $p$-values of six tests over $2000$ replications under each of the following three fixed null hypotheses with $H_{\Omega_{0, 1}}: (\alpha_S, \beta_M) = (0.5, 0)$, $H_{\Omega_{0, 2}}: (\alpha_S, \beta_M) = (0, 0.5)$,  and $H_{\Omega_{0, 3}}: (\alpha_S, \beta_M) = (0, 0)$. 
\Cref{fig:QQ_plot} shows Q-Q plots, which are expected to be uniformly distributed when type I error is properly controlled. 
It is evident that in the scenarios of $H_{\Omega_{0,1}}$ and $H_{\Omega_{0,2}}$, all tests' $p$-value distributions are approximately uniformly distributed over $(0,1)$ except for PoC-B and PoC-YM, due to the violation of the linearity and/or normality assumptions by these two tests under the generalized SEM. 
It is interesting to note that none of the competing methods, including the naive nonparametric bootstrap test QMA-B can properly control the type I error under $\Omega_{0, 3}$. {Unfortunately, this is the method that has been taken for granted in practice. } Our proposed method is the only AB test that produces uniformly distributed $p$-values under all three cases. This desirable type I error control leads to power gain as seen in \Cref{section:simulation:power}.

\begin{figure}[htp]
\centerline{\renewcommand{\arraystretch}{0.8} 
	\begin{tabular}{ccc}			
		$H_{\Omega_{0,1}}: (\alpha_S, \beta_M) = (0.5, 0)$  & $H_{\Omega_{0,2}}: (\alpha_S, \beta_M) = (0, 0.5)$   & $H_{\Omega_{0,3}}: (\alpha_S, \beta_M) = (0, 0)$ \\[-2ex] \psfig{figure=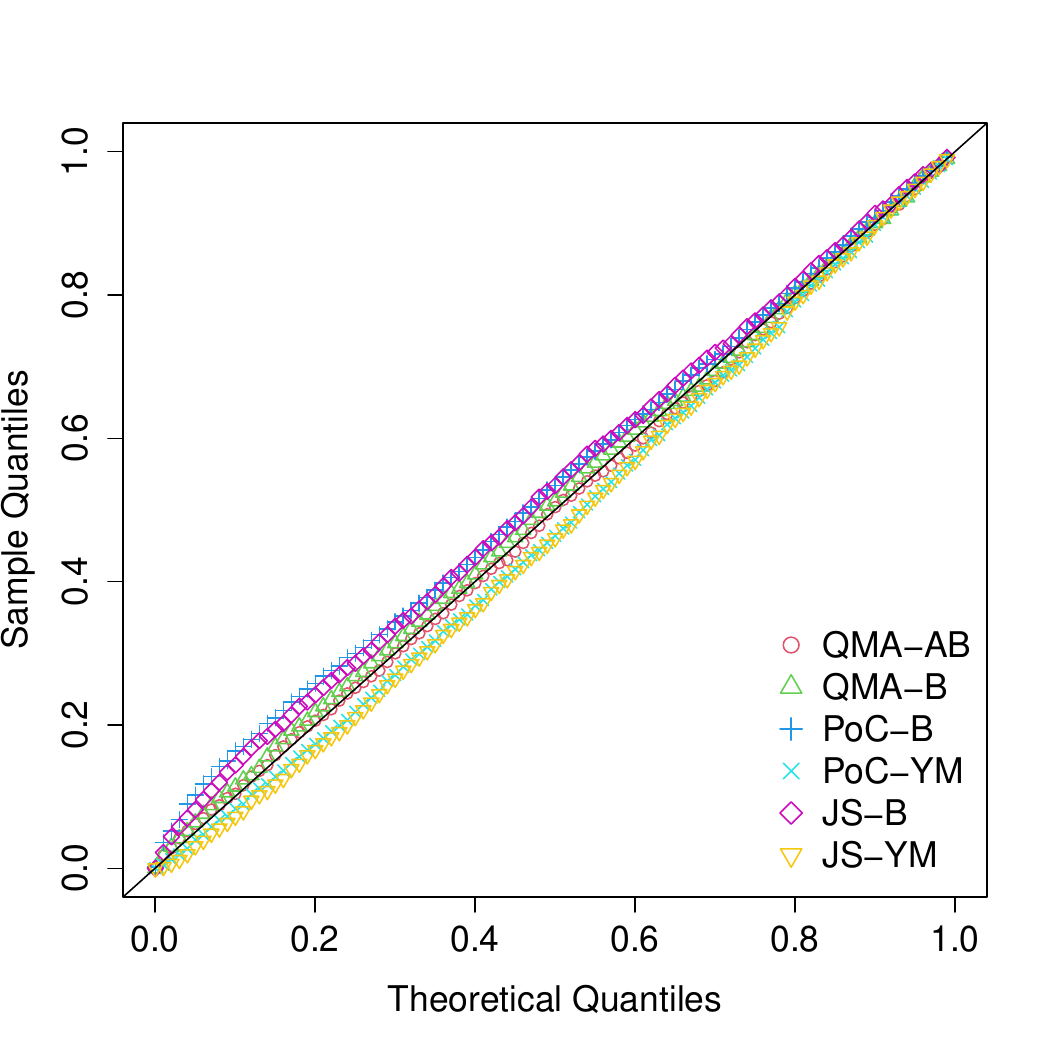,width=1.8in,angle=0} & \psfig{figure=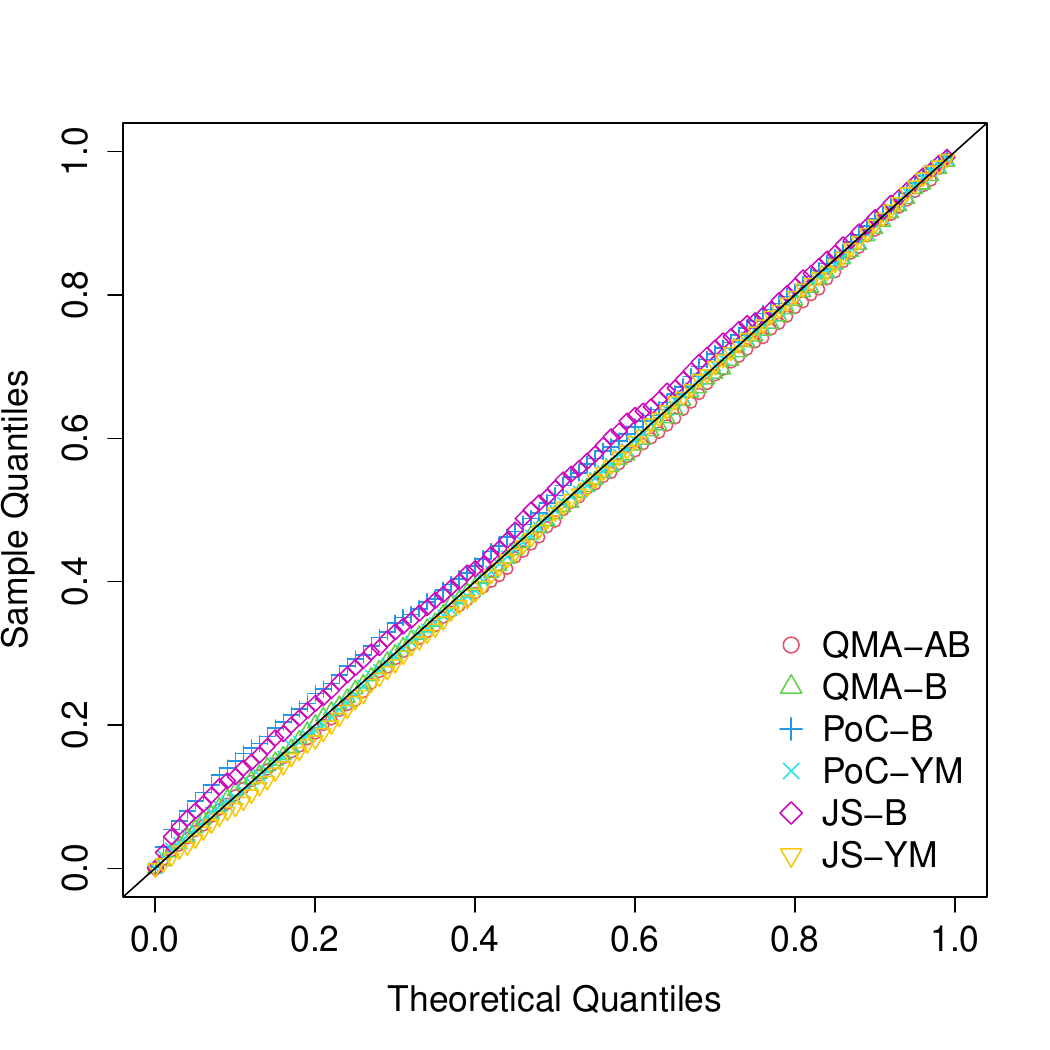,width=1.8in,angle=0} & \psfig{figure=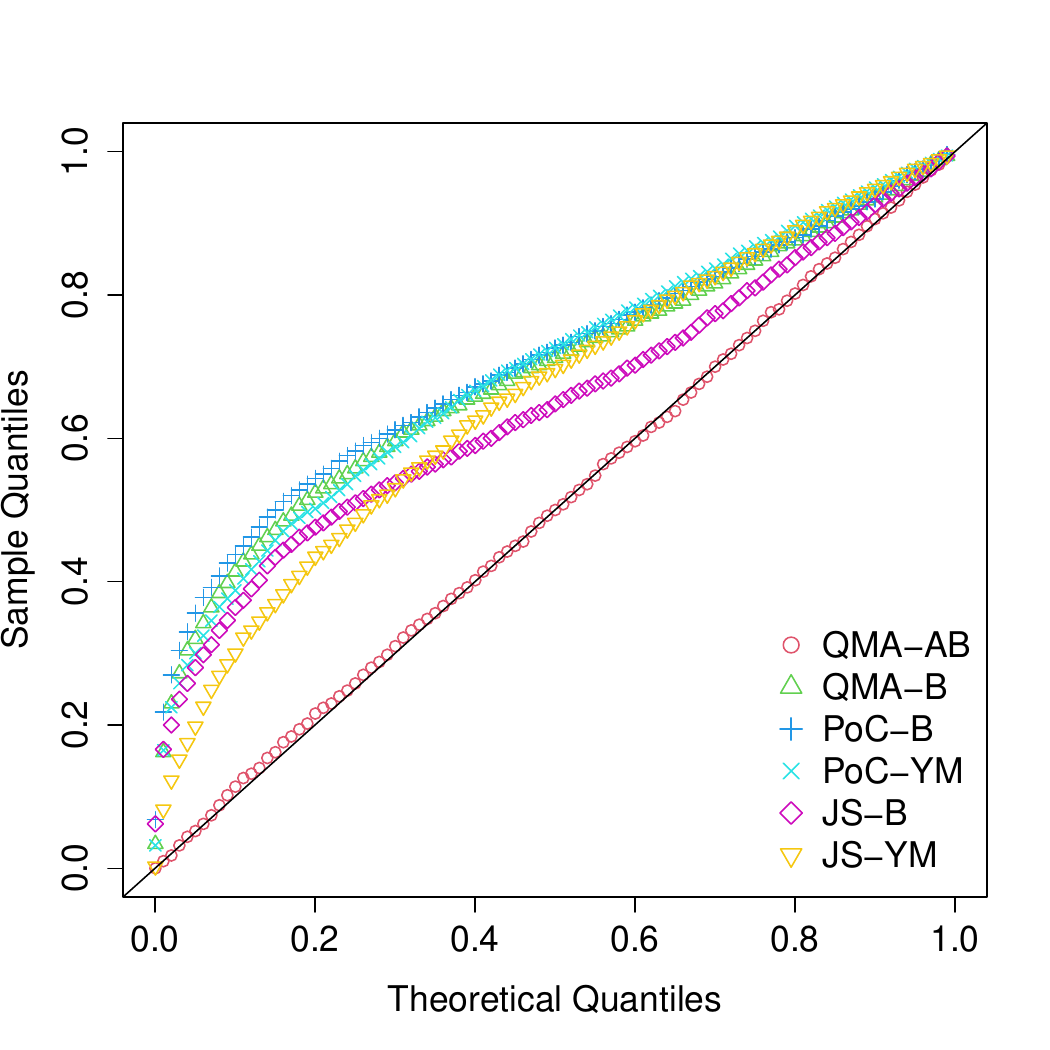,width=1.8in,angle=0}\\
	\end{tabular}
}
\captionsetup{font = footnotesize}\vskip-0.6cm
\caption{Q-Q plots of $p$-values under the three cases of fixed null hypotheses with sample size $n = 300$.}
\label{fig:QQ_plot}
\end{figure}

{\it Setting II: A mixture of nulls.} In this simulation design,  we vary the configuration of the null hypotheses. That is, at each replication, a null is randomly drawn from $H_{\Omega_{0, 1}}$, $H_{\Omega_{0, 2}}$ and $H_{\Omega_{0, 3}}$ defined above in Setting I. We consider three selection probabilities: (A) $(1/3, 1/3, 1/3)$, (B) $(0.2, 0.2, 0.6)$, and (C) $(0.05, 0.05, 0.9)$, in which the proportion of $\Omega_{0,3}$ increases from scenario (A) to scenario (C). This design mimics real-world applications where the test is repeatedly used for a large number of mediators that may have different mediation pathways.  \Cref{fig:QQ_plot_mixture} displays Q-Q plots of $p$-values under these three selection probabilities for non-fixed null hypotheses. 

{Interestingly,} our proposed AB test performs steadily well under all the settings, giving rise to high confidence of the methodology to be used in practice.  
It is easy to visualize that all five competing tests exhibit excessive conservation in all scenarios (A)-(C), and become increasingly conservative as the proportion of $\Omega_{0, 3}$ increases. Arguably, this may pose a concern in biological studies with many mediators, especially where null subspace $\Omega_{0, 3}$ predominates the configuration of null cases. Therefore, our AB test may enhance scientific discoveries with proper type I error control and desirable power.

\begin{figure}[htp!]
\centerline{\renewcommand{\arraystretch}{0.8} 
	\begin{tabular}{ccc}		A: $(1/3, 1/3, 1/3)$& B: $(0.2, 0.2, 0.6)$&  C: $(0.05, 0.05, 0.9)$ \\[-2ex]	
		\psfig{figure=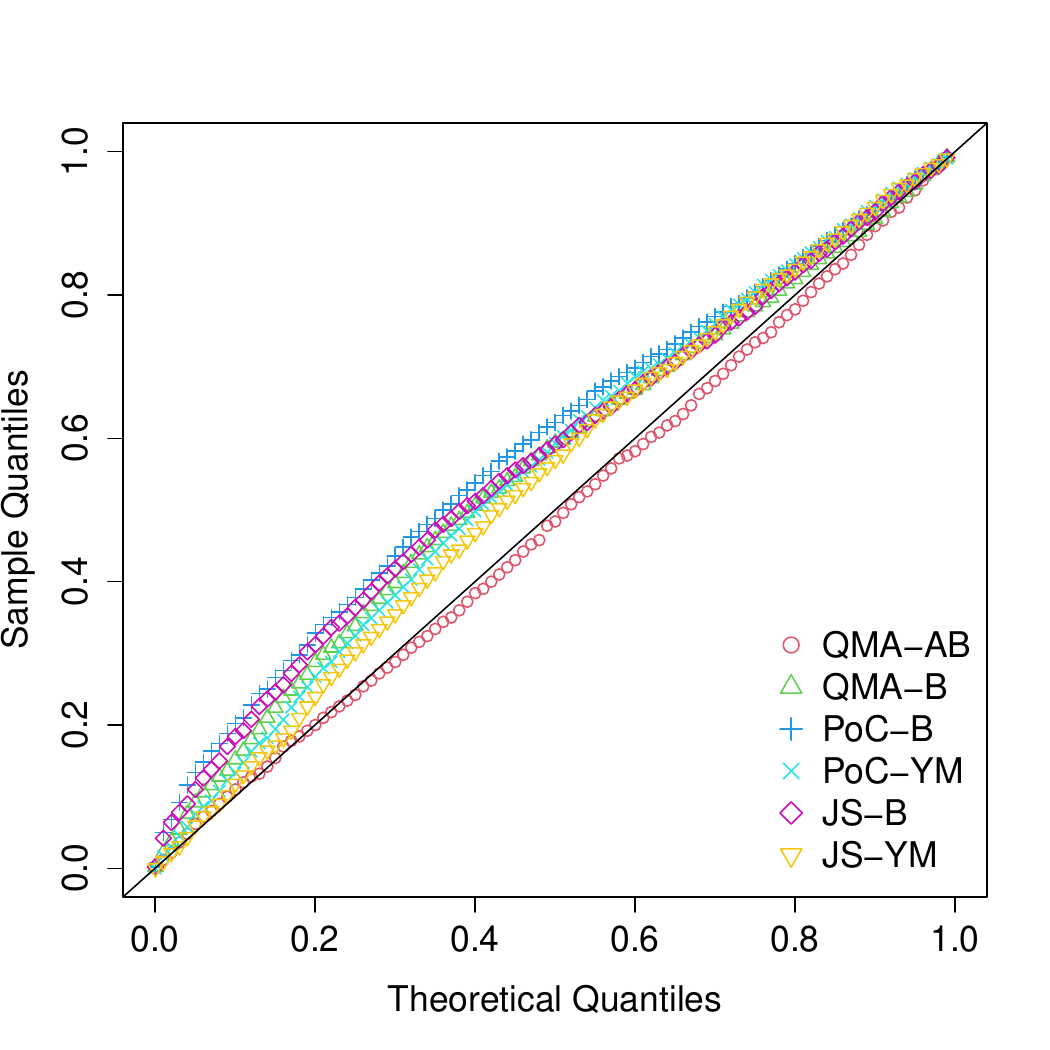,width=1.8in,angle=0} & \psfig{figure=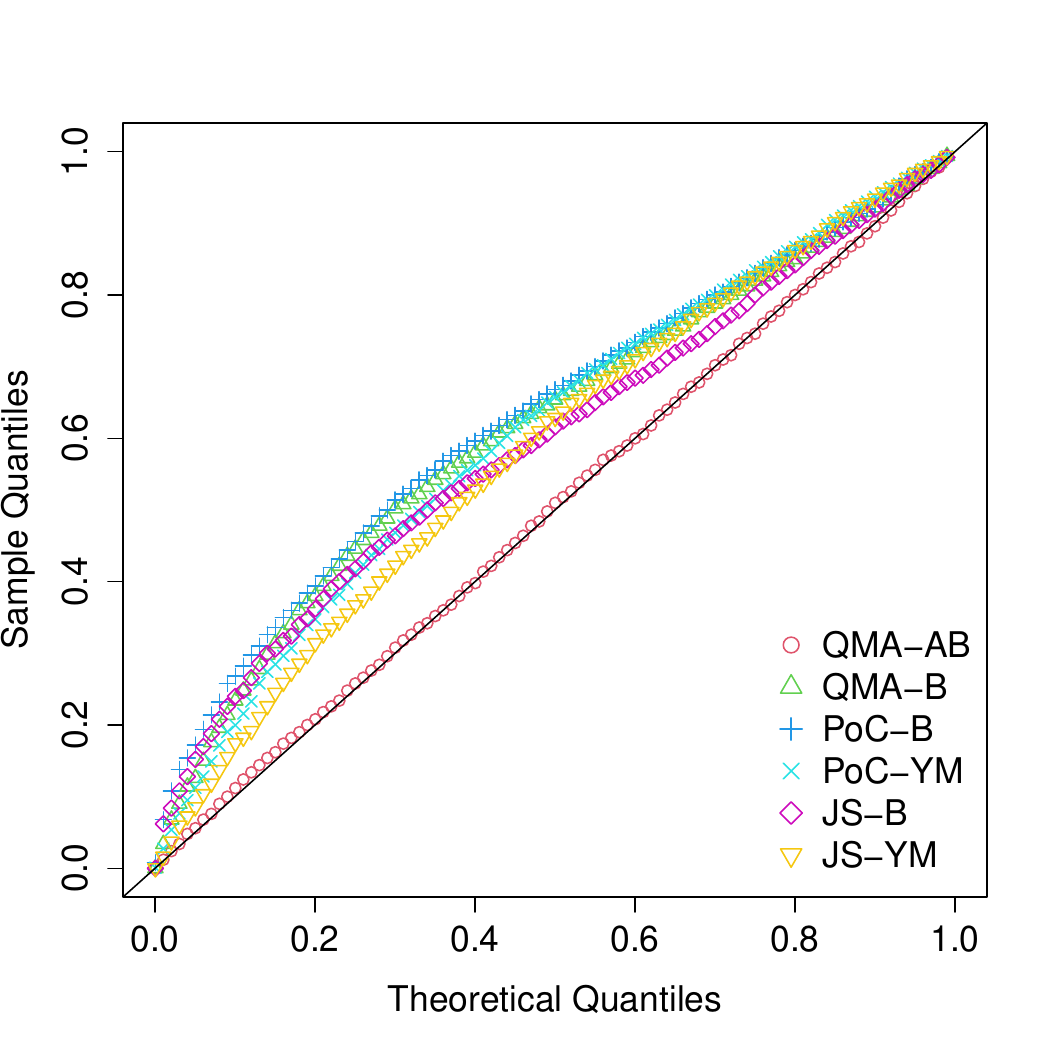,width=1.8in,angle=0} & \psfig{figure=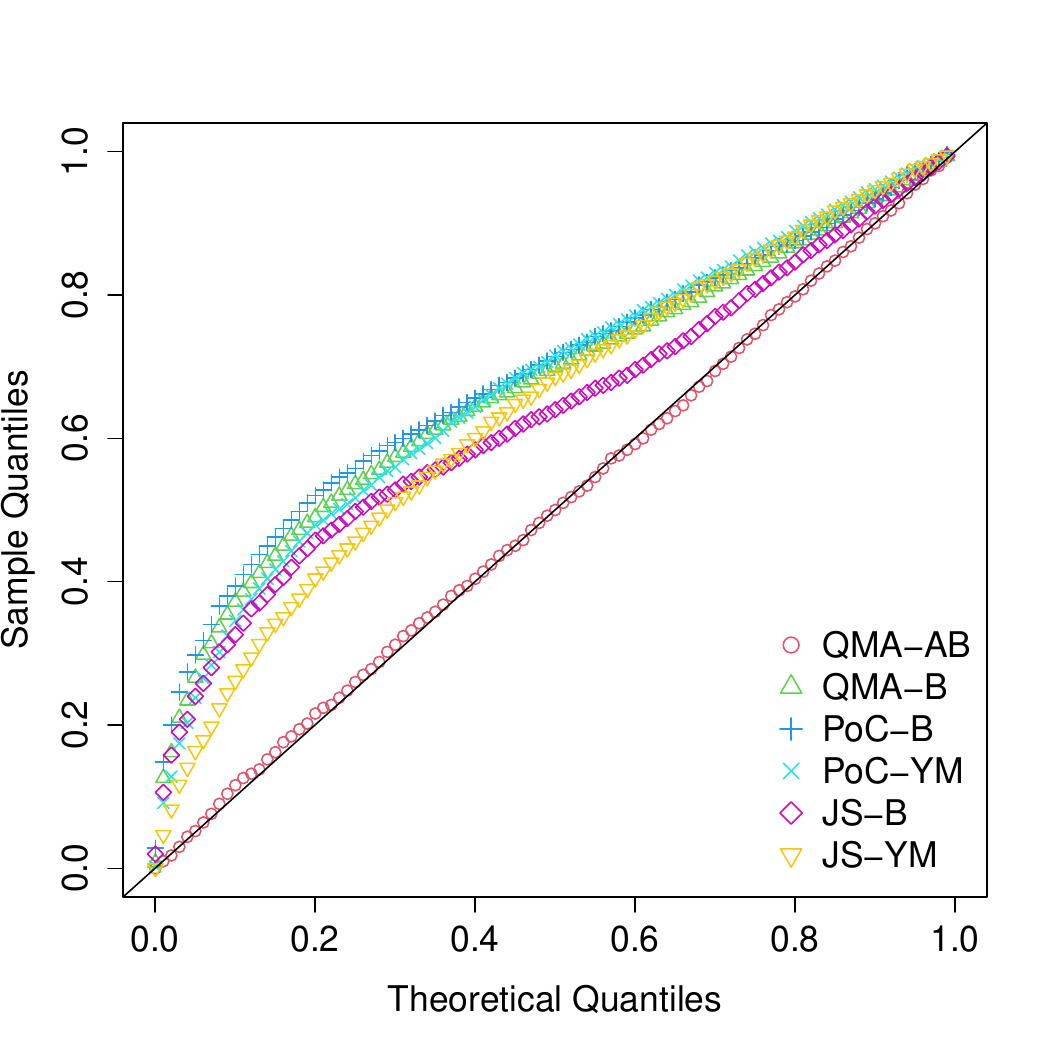,width=1.8in,angle=0}\\
	\end{tabular}
}
\captionsetup{font = footnotesize}\vskip-0.6cm
\caption{Q-Q plots of $p$-values under the non-fixed null hypotheses with these selection probabilities and sample size $n = 300$. }
\label{fig:QQ_plot_mixture}
\end{figure}

\subsection{Power evaluation}\label{section:simulation:power}
We assess the statistical power of our proposed AB test in two settings: (i)  $\alpha_S = \beta_M$ that gradually increases from 0 to 0.2, and (ii) $\alpha_S\beta_M = 0.2^2$ with varying ratio of $\alpha_S/\beta_M$.
In \Cref{fig:power_I}, we present the empirical rejection rates over $2000$ replications against the signal size $\alpha_S$ (the left panel) and the ratio (the right panel), respectively. With no surprise, the best type I error control gives the highest return for the statistical power, justified by the evidence that our QMA-AB exhibits the largest power curves than the five competing methods.

{\color{blue}
\begin{figure}[htp!]
	\centerline{
		\begin{tabular}{cc}		
			$\alpha_S = \beta_M$& $\alpha_S/\beta_M$ \\[-4ex]	
			\psfig{figure=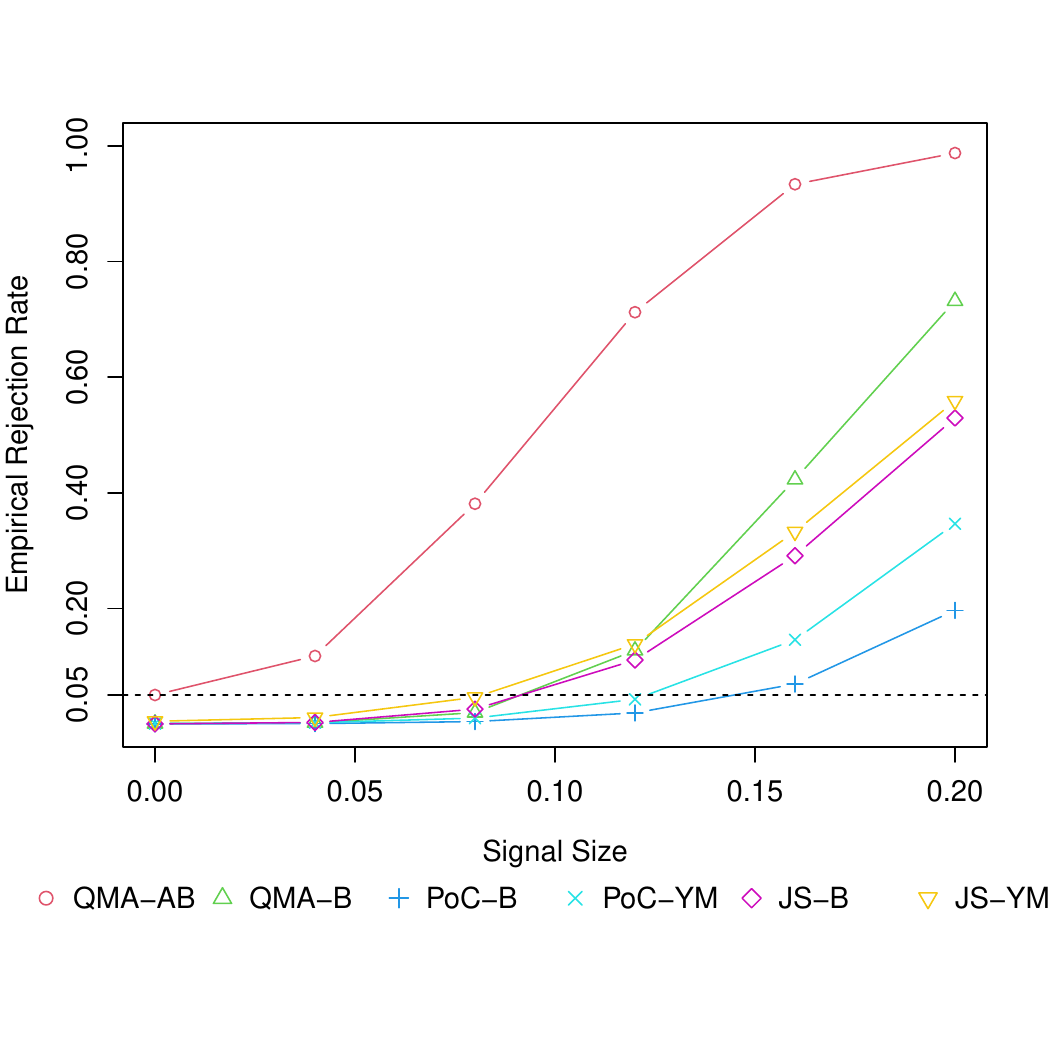,width=2.3in,angle=0} 
			& \psfig{figure=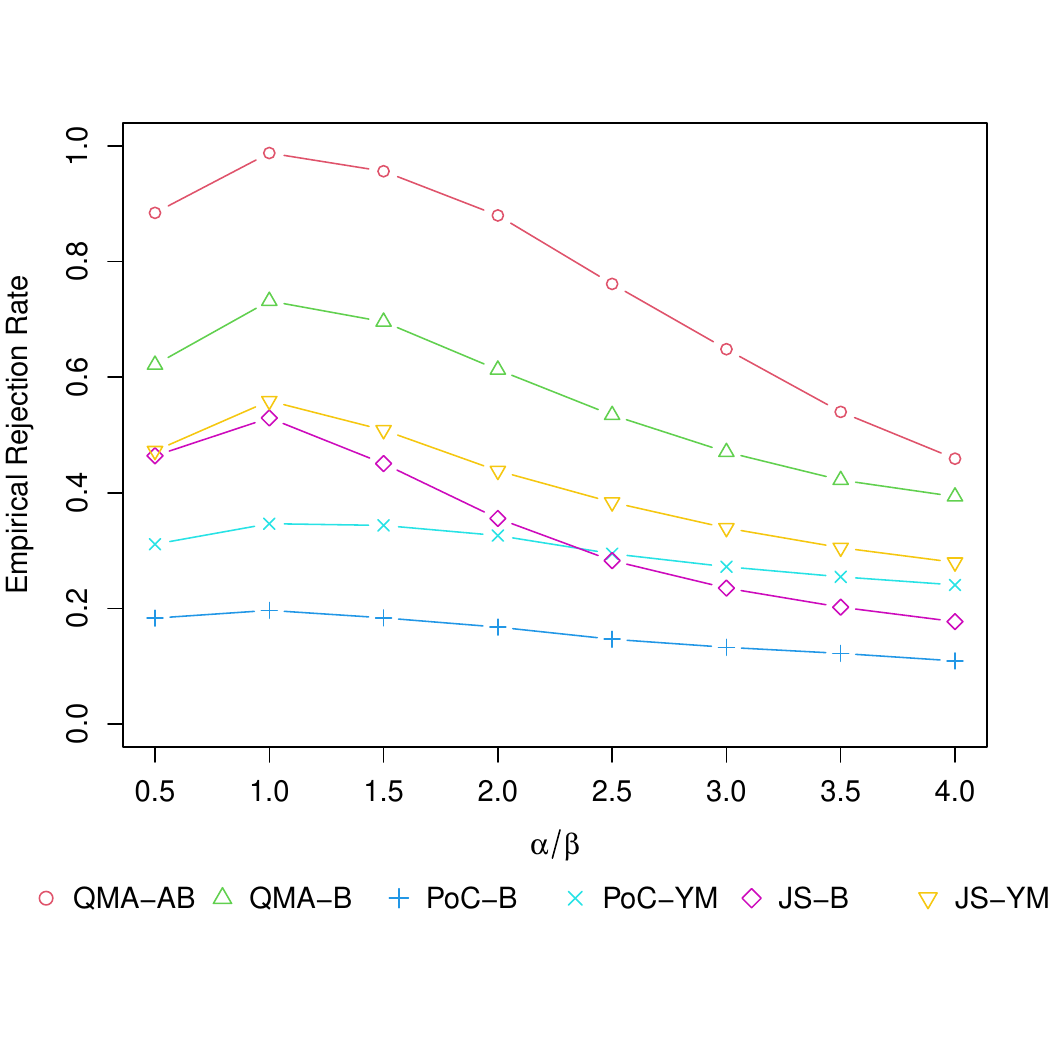,width=2.3in,angle=0}\\
		\end{tabular}
	}
	\captionsetup{font = footnotesize}\vskip-1.2cm
	\caption{Empirical rejection rate of the six tests over signal strength of the mediation pathway. }
	\label{fig:power_I}
\end{figure}
}

\section{Childhood Obesity Metabolic Study}\label{section:real_data}
We apply our proposed methodology to analyze a dataset collected from the ELEMENT cohort study \citep{perng2019EarlyLifeExposure}. The dataset comprises 291 adolescents aged 10--18 years from Mexico City. The pressing need for potential intervention options against the early onset of obesity 
has drawn significant attention to environmental toxicants, such as phthalates, which have been identified as ubiquitous risk factors for childhood obesity \citep{yang2017BisphenolPhthalatesUtero}. Our data analysis aims to identify potential mediation pathways in that metabolites, particularly lipids, may mediate the association of exposure to toxicants with childhood obesity. Molecular-level alterations in metabolic profiles triggered by certain toxicants may facilitate the discovery of important mediation pathways associated with early onset of obesity.

We focus on a target toxicant, maternal phthalate MEOHP, measured in the 3rd trimester of pregnancy, a ubiquitous chemical existing in food production and storage. Childhood obesity, defined by a BMI at or above the 95th percentile for children and teens of the same age and sex \citep{centersfordiseasecontrolandprevention2022ChildTeenBMI}, points to the choice of $\tau = 0.95$ in our analysis.  We focus on 158 lipid metabolites as potential mediators, and aim to identify important mediation pathways of $\text{MEOHP}\to\text{lipid}\to\text{childhood obesity}$ at the quantile level $\tau = 0.95$. Eight potential confounders are included: mother's age at delivery, birth weight, number of pregnancies, gestational age, education, marital status, and child's age and sex. Following the literature, we transform MEOHP using the square root, and lipids are all transformed into z-scores in the analysis. According to \cite{juvanhol2016FactorsAssociatedOverweight}, BMI is modeled by a Gamma distribution in the generalized SEM.

We apply six methods that have been compared in the simulation studies: QMA-AB, QMA-B, PoC-B, PoC-YM, JS-B, and JS-YM, with $2000$ bootstrap samples. For ease of interpretation, we set reference confounders as girl of 16.38 years with a birth weight of 3.13 kg, a gestational age of 38.73 weeks, whose mothers are married, aged 26.85 years old, have received 10.78 years of education, and have had 2 pregnancies. Exposure to MEOHP changes from zero (no exposure) to a one-unit increase (i.e., $s = 0$ to $s^\prime = 1$).

First, we assess the overall set-level significance of the 158 lipids using the Cauchy combination test \citep{liu2020CauchyCombinationTest}, and find $p$-values equal to 0.0375 (QMA-AB), 0.7339 (QMA-B), 0.8029 (PoC-B), 0.8722 (PoC-YM), 0.7805 (JS-B), and 0.3574 (JS-YM. Of note, only our AB test detected the overall statistical significance for the mediation pathway $\text{MEOHP}\to \text{lipid set}\to \text{childhood obesity}$, where the other five methods fail to detect any signals.

Second, we examine individual-level pathways, one lipid at a time. Significant pathways are selected based on $p$-values obtained by QMA-AB, adjusting for multiple comparisons with a controlled false discovery rate (FDR) at level $0.1$ \citep{benjamini1995ControllingFalseDiscovery}. We identify seven significant lipids that are significant mediators for the pathway of $\text{MEOHP}\to \text{lipid}\to \text{childhood obesity}$; see \Cref{table:real_data}. {The estimates of $\QNDE$, $\QTE$, $\alpha_S$,  $\beta_M$ and $\gamma_S$ are listed in \Cref{table:real_data:estimated_alpha_beta} of the Supplementary Material.}

\begin{table}[ht]
\caption{Seven significant lipids found in the quantile mediation analysis of $\text{MEOHP}\to \text{lipid}\to \text{childhood obesity}$ at controlled FDR rate of $0.1$. }\footnotesize
\label{table:real_data}
\centering
\begin{tabular}{lccccccc}
	& & \multicolumn{6}{c}{$p$-value}\\
	\cline{3-8} 
	Lipid & $\QNIE_{0.95}$  & QMA-AB & QMA-B & PoC-B & PoC-YM & JS-B & JS-YM  \\
	MG 14:0 & -0.3168 & 0.0010 & 0.0880 & 0.4095 & 0.3475 & 0.7015 & 0.3054 \\
	PC 38:2/PE 41:2 & -0.3438 & 0.0020 & 0.0955 & 0.1130 & 0.0833 & 0.5020 & 0.0154 \\
	FA 26:1 DiC & -0.4475  & 0.0025 & 0.1110 & 0.8985 & 0.8971 & 0.8375 & 0.8946 \\
	FA 15:0 DiC &  0.4504 & 0.0025 & 0.0655 & 0.7410 & 0.7170 & 0.7655 & 0.7143 \\
	PC 34:3/PE 37:3 & -0.3803 & 0.0025 & 0.1000 & 0.7835 & 0.7613 & 0.8125 & 0.7592 \\ 
	PC 32:2/PE 35:2 & -0.3534 & 0.0035 & 0.1360 & 0.1715 & 0.0441 & 0.3445 & 0.0069 \\
	PC 40:4/PE 43:4 & -0.3534 & 0.0040 & 0.0355 & 0.0945 & 0.0116 & 0.5600 & 0.0018 \\
\end{tabular}
\end{table}

The identified lipids offer valuable insights into the intricate metabolic network associated with lipid mediators and their implications for obesity-related pathways.
{A summary for the topmost signal MG 14:0 in \Cref{table:real_data} is given below, and more details can be found in \Cref{section:additional_numerical_results:real_data} of  the Supplementary Material.}
Lipid MG 14:0, characterized as a diacylglycerol and a triacylglycerol (TAG) precursor, is {responsible} for decreasing serum TAG concentration, TAG synthesis, and increasing energy expenditure, potentially reducing fat accumulation \citep{yuan2010DiacylglycerolOilReduces}. Thus, MG 14:0 may be considered as a prospective peripheral target for new anti-obesity therapeutics \citep{take2016PharmacologicalInhibitionMonoacylglycerol}. %

{
Closing the analysis, we conducted a sensitivity analysis as well as model diagnosis to evaluate both validity of sequential ignorability and goodness-of-fit of the generalized SEM. Our sensitivity analysis follows the procedure proposed by \cite{imai2010GeneralApproachCausal}, \cite{liu2022LargeScaleHypothesisTestinga} and \cite{hao2023ClassDirectedAcyclic}. First, we take normal inverse transformations for the triplet $(S, M, Y)$ in the generalized SEM \eqref{model:GSEM}: $Z_S = \Phi^{-1}\{F_{S\mid \X}(S\mid \x)\}$, $Z_M = \Phi^{-1}\{F_{M\mid \X}(M\mid \x)\}$ and $Z_Y = \Phi^{-1}\{F_{Y\mid \X}(Y\mid \x)\}$. Then, conditioning on $\X = \x$,  
$
Z_S\sim\calN(0, 1),\enskip Z_M\mid Z_S\sim \calN(\alpha_S Z_S, 1)$, and $Z_Y\mid Z_S, Z_M\sim \calN(\gamma_S Z_S + \beta_M \delta_M Z_M, 1)$, where $\delta_M = (\alpha_S^2 + 1)^{1/2}$.  
Let $\varepsilon_M = Z_M - \alpha_S Z_S$ and $\varepsilon_Y = Z_Y - \gamma_S Z_S + \beta_M \delta_M Z_M$. In the generalized SEM, errors $\varepsilon_M$ and $\varepsilon_Y$ are independently normally distributed. When the sequential ignorability condition is violated, it is likely that the correlation $\rho = \text{corr}(\varepsilon_M, \varepsilon_Y)\neq 0$ and {\it vice versa}. Following \cite{imai2010GeneralApproachCausal}, we hypothetically vary the value of $\rho$ and compute the corresponding $\wh\QNIE_{0.95}$. For each lipid, we compute $\abs{\rho_{b}}$, the breakpoint of $\abs{\rho}$ such that the estimate $\wh\QNIE_{0.95} = 0$. All lipids with $\abs{\wh\QNIE_{0.95}}>0.2$ (at $\rho = 0)$ are plotted in \Cref{fig:real_data:sensitivity:model_checking} (a). For the top mediator MG 14:0, \Cref{fig:real_data:sensitivity:model_checking} (a) indicates that when $\text{corr}(\varepsilon_M, \varepsilon_Y)>0.11$ in order for $\wh\QNIE_{0.95}=0$. A simple calculation unveils that the absolute sample correlation for MG 14:0 is 0.0007, which is much smaller than $0.11$. This suggests that our discovery of MG 14:0 seems trustworthy, which holds the ground for certain mild violations from the zero correlation. Similar findings are obtained for the rest mediators. Thus, the sensitivity analysis shows that our analysis for the ELEMENT dataset is trustworthy.

Finally, we conduct model diagnosis on the generalized SEM in the analysis. We adopt the ``in-and-out-of-sample'' likelihood ratio test proposed by \cite{zhang2016GoodnessoffitTestSpecification} to flag potential model misspecification.  All $p$-values of the test are plotted in   \Cref{fig:real_data:sensitivity:model_checking} (b). At 0.1 FDR, we have identified only two lipids, FA 10:0 and FA 8:0, that fail to pass the goodness-of-fit test. However, these two lipids are not in the pool of significant mediators. Nevertheless, we shall always be cautious when making any conclusions on such mediators in a model-based analysis.

}
\begin{figure}[htbp!]
\centerline{\renewcommand{\arraystretch}{0.8} 
	\begin{tabular}{cc}	
		\psfig{figure=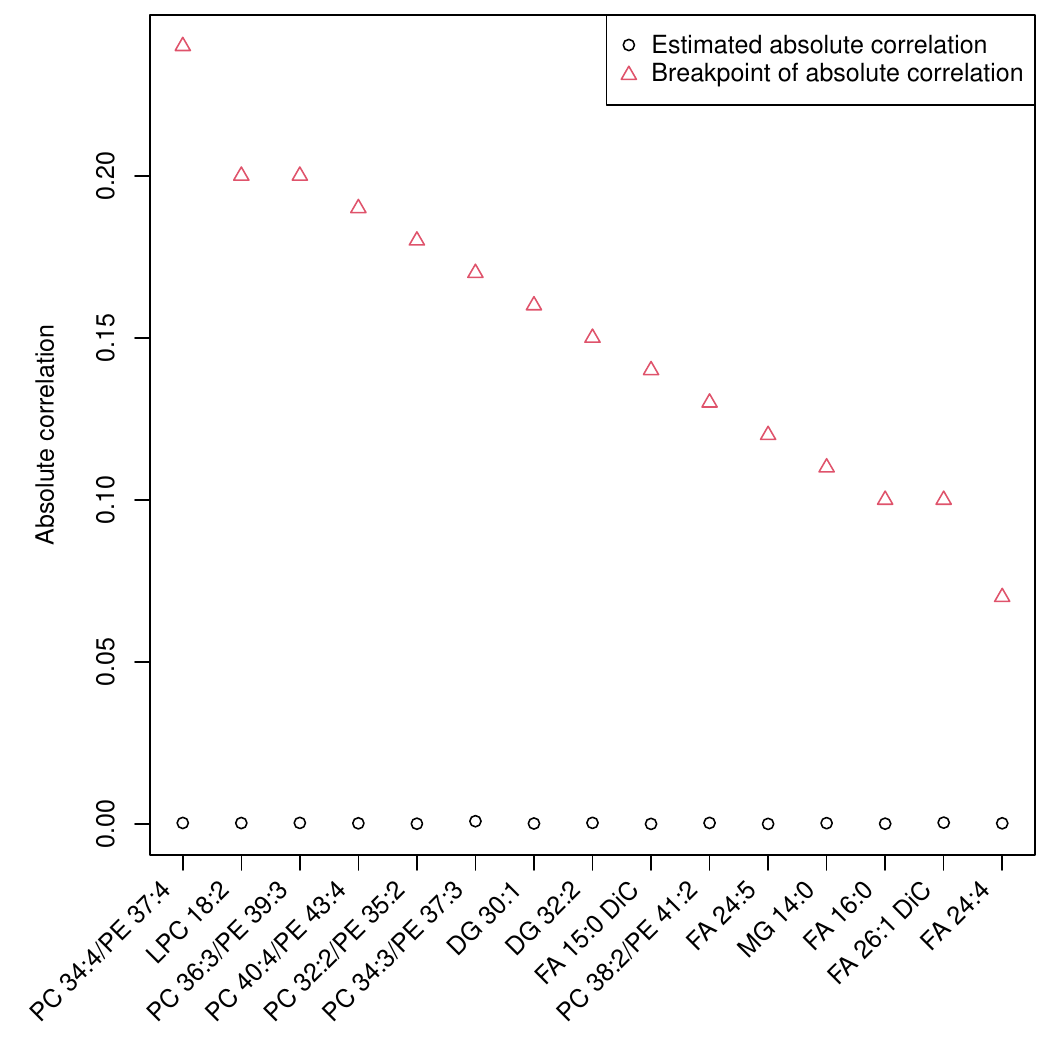,width=2.2in,angle=0} & \psfig{figure=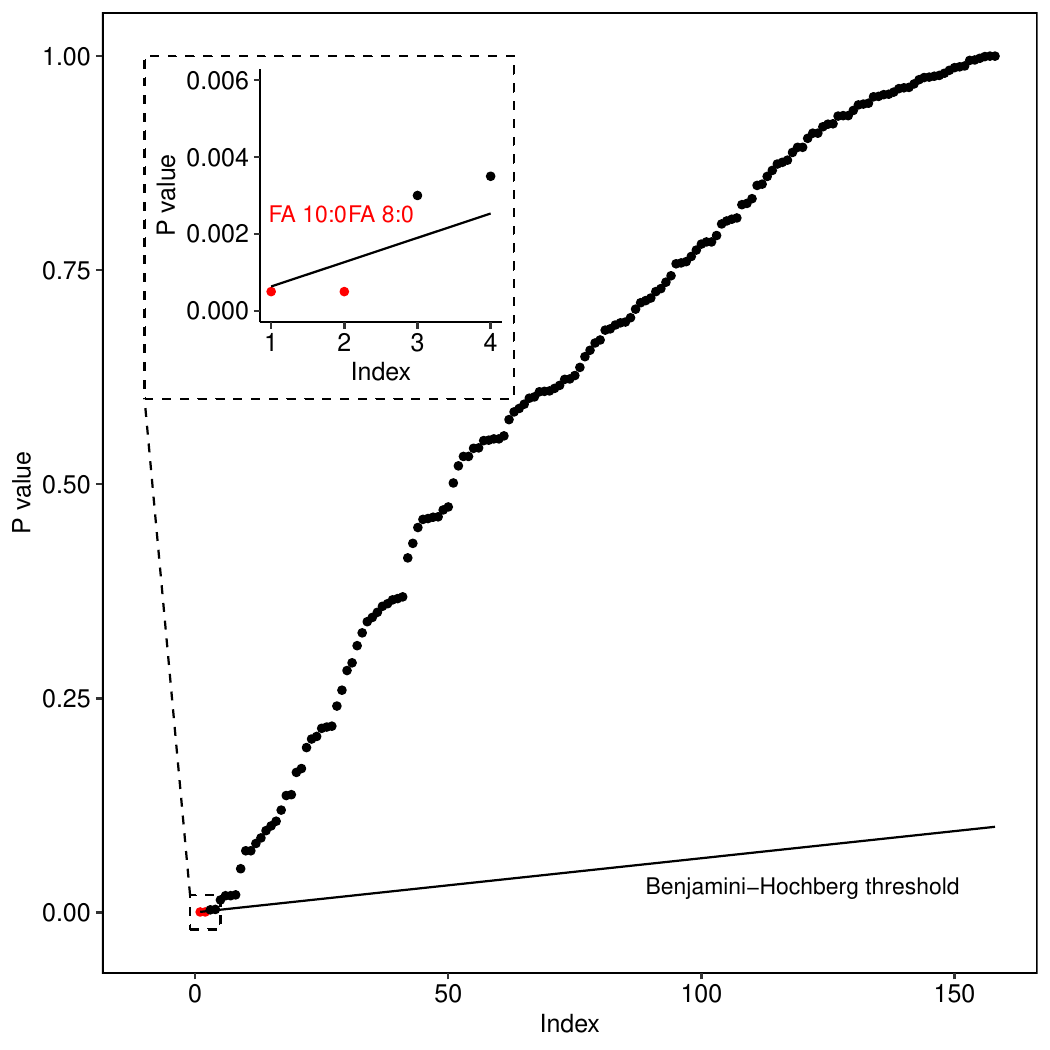,width=2.2in,angle=0} \\
		(a) Sensitivity Analysis & (b) Model Diagnosis
	\end{tabular} 
}\vskip-0.4cm
\captionsetup{font = footnotesize}
\caption{(a) The estimated breakpoint of the absolute correlation at which $\QNIE_{0.95} = 0$ among lipids whose $\abs{\wh\QNIE_{0.95}}>0.2$, and the estimated absolute correlation. 
	(b) The $p$-values of the goodness-of-fit test for 158 lipids in anscending order.}
\label{fig:real_data:sensitivity:model_checking}
\end{figure}

\section{Concluding Remarks}\label{section:conclusion}

We developed a new methodology to analyze quantile mediation pathways in the context of generalized SEMs. We discussed in detail both the identification condition and sequential ignorability assumption under which we derived closed-form expressions for quantile mediation estimands. The main technical contribution lies in the development of an adaptive bootstrap procedure that can discern different null subspaces in the construction of test statistics. We have established key theoretical properties, including the bootstrap consistency for the proposed testing method. Extensive simulations have confirmed the theoretical properties, clearly showing that our AB test for the mediation effect has a properly controlled type I error, which leads to significantly improved statistical power over existing methods. One surprise is that only two of 158 lipids may be not modeled by the generalized SEM according to the goodness-of-fit test. This gives the confidence of breadth and flexibility of the generalized SEM.

Our new methodology allows us to consider several important extensions. One is to extend the generalized SEM with the high-dimensional covariates  \citep{guo2022HighDimensionalMediationAnalysisa}. In practice, practitioners may include many covariates to increase the chance of making the sequential ignorability assumption a valid condition. The proposed generalized SEM is structured by hierarchical modeling layers that permit the inclusion of high-dimensional covariates in all marginal GLMs. The second extension deals directly with a set of mediators \citep{he2023AdaptiveBootstrapTests,hao2024SimultaneousLikelihoodTest}, rather than the use of Cauchy combination test. Currently, little work has been done for a joint quantile mediation analysis involving a set of mediators.  Flexibility of the generalized SEM may allow us to handle multi-mediators, which is worth further investigation.

\section*{Acknowledgement}
This research was partially funded by NIH R01ES033565. 


\section*{Supplementary material}
\label{SM}
Supplementary Material includes proofs for all theoretical results in the main text, additional simulation results and interpretations for the identified signals. 




\appendix
\renewcommand{\thesection}{\Alph{section}}

\renewcommand{\theequation}{\Alph{section}.\arabic{equation}}
\renewcommand{\theremark}{\Alph{section}.\arabic{remark}}
\renewcommand{\thefigure}{\Alph{section}.\arabic{figure}}
\renewcommand{\thetable}{\Alph{section}.\arabic{table}}

\setcounter{section}{0} 
\setcounter{figure}{0} 
\setcounter{table}{0} 
\setcounter{remark}{0} 

\section{Generalized Structural Equation Model}\label{section:generalized_SEM}

A DAG, shown in~\Cref{fig:mediationDAG_unconfounded} (A), involves three random variables including an exposure $S\in\mR^1$, a mediator $M\in\mR^1$, and an outcome $Y\in\mR^1$. A classical SEM for this DAG takes the form: $S  =  \varepsilon_S$, $
M  =  \alpha_S S + \varepsilon_M$, and $
Y  =  \beta_M M + \gamma_S S + \varepsilon_Y$. This gives an algebraic presentation of graphic topology  \citep{pearl2014InterpretationIdentificationCausal}. With the covariance matrix of the three error terms $(\varepsilon_S, \varepsilon_M,\varepsilon_Y)^T$, $\bSig\defn \diag(\sigma^2_{\varepsilon_S}, \sigma^2_{\varepsilon_M}, \sigma^2_{\varepsilon_Y})$, and the {\it weighted adjacency matrix}
\beqrs
\bTh = 
\begin{pmatrix}
0 & 0 & 0 \\
\alpha_S & 0 & 0\\
\gamma_S & \beta_M & 0\\
\end{pmatrix} \defn \LT(\alpha_S, \gamma_S, \beta_M),
\eeqrs
clearly, the covariance of $(S, M, Y)$ is 
\beqrs
\bGam\defn \var\{(S, M, Y)\trans\} =  (\I - \bTh)^{-1}\bSig(\I - \bTh)^{-{\mbox{\tiny{T}}}}. 
\eeqrs%
Thus parameters
$(\alpha_S, \gamma_S, \beta_M)$ appear {solely in the second moment} of the joint distribution. As shown in  \Cref{fig:mediationDAG_unconfounded} (B), the classical SEM adds confounding factors $\X\in\mR^p$ as follows: {$S  =  \bbeta_S\trans\X + \varepsilon_S$, $M  =  \alpha_S S + \bbeta_M\trans\X + \varepsilon_M$, and $ Y  =  \beta_M M + \gamma_S S + \bbeta_Y\trans\X + \varepsilon_Y$. The confounding enters the SEM via linear predictors (e.g. $\bbeta_S\trans\X$) that affect {only the first moment of the joint distribution};  see \Cref{appendix:hierarchical_modelling} of the Supplementary Material for details.} Such SEM introduces a space perpendicular to $\mbox{span}(\X)$ so that {removal} of the linear predictors would have no impact on the DAG topology. That is, the classical SEM suggests that the linear predictors of $\X$ and DAG topology $\bTh$ reside in two orthogonal spaces. These insights are incorporated into the following extension proposed by \cite{hao2023ClassDirectedAcyclic} termed generalized SEM. 

\cite{hao2023ClassDirectedAcyclic} suggested incorporating the covariance structure $\bGam$ 
in the Gaussian copula \citep{song2000MultivariateDispersionModels}, $C(u_1, u_2, u_3) = \Phi_3\{\Phi^{-1}(u_1), \Phi^{-1}(u_2), \Phi^{-1}(u_3); \bGam^\prime\}$, where $\Phi_3(\cdot; \bGam^\prime)$ is a trivariate Gaussian distribution function with mean zero and correlation matrix $\bGam^\prime$, and $\Phi(\cdot)$ is the standard Gaussian distribution function. 
In this hierarchical modeling with copula, all the marginal parameters including both {regression} parameters and variance parameters are included in the marginal distributions, where matrix $\bGam = (\I - \bTh)^{-1}(\I - \bTh)^{-{\mbox{\tiny{T}}}}$ contains only correlation parameters. The resulting joint distribution is expressed as 
\beqrs
F_{S, M, Y\mid \X}(s, m, y\mid \x) = C\{F_{S\mid \X}(s\mid \x), F_{M\mid \X}(m\mid \x), F_{Y\mid \X}(y\mid \x); \bGam^\prime\}.
\eeqrs
Let $\eta = \alpha_S\beta_M + \gamma_S$ be the total DAG effect. The induced correlation matrix $\bGam^\prime$ in the copula joint model has an explicit form:
\beqr
\label{model:bGam}
\bGam^\prime = \left(\begin{array}{ccc}1 & \frac{\alpha_S}{(\alpha_S^2+1)^{1/2}} & \frac{\eta}{(\eta^2+\beta_M^2+1)^{1/2}} \\ \frac{\alpha_S}{(\alpha_S^2+1)^{1/2}} & 1 & \frac{\alpha_S \eta+\beta_M}{(\alpha_S^2+1)^{1/2} (\eta^2+\beta_M^2+1)^{1/2}} \\ \frac{\eta}{(\eta^2+\beta_M^2+1)^{1/2}} & \frac{\alpha_S \eta+\beta_M}{(\alpha_S^2+1)^{1/2} (\eta^2+\beta_M^2+1)^{1/2}} & 1\end{array}\right).
\eeqr
It is interesting to note that the dependence between exposure $S$ and outcome $Y$ is $\mbox{corr}(S, Y) = {\eta}/{(\eta^2+\beta_M^2+1)^{1/2}}$, a term that contributes to the drift along with a given quantile level in \Cref{theorem:explicit_form_continuous_case}.  
\begin{remark}
The generalized SEM extends the classical linear SEM by allowing non-linear dependence among variables $(S, M, Y)$. Because $\Phi^{-1}\{F(\cdot)\}$ is a non-linear monotone transformation, the correlations in $\bGam^\prime$ have close connections with rank-based correlation measures such as Spearman's $\rho$, Kendall's $\tau$, and Chatterjee's $\xi$ \citep{zhang2022SlicedIndependenceTest}.  When $F$ is normal, the generalized SEM reduces to the classical linear SEM {along with Pearson correlation.}
\end{remark}

We extend the marginal linear regression to the generalized linear model (GLM) for modeling the marginal distributions. Specifically, we assume that given the covariates $\X$, the marginal distribution of $S$, $M$, and $Y$ are elements of exponential dispersion family distribution (ED) models {\citep{jorgensen1987ExponentialDispersionModels}}, that is, 
\beqrs
& S\mid \X \sim \text{ED}(\mu_S(\X), \phi_S),  & g_X(\mu_S(\X)) = \X\trans\bbeta_S,\enskip  M\mid \X \sim \text{ED}(\mu_M(\X), \phi_M),  g_M(\mu_M(\X)) = \X\trans\bbeta_M\\
& Y\mid \X \sim \text{ED}(\mu_Y(\X), \phi_Y),  & g_Y(\mu_Y(\X)) = \X\trans\bbeta_Y,
\eeqrs
where $g_X$, $g_M$ and $g_Y$ are known link functions, and $\phi_S$, $\phi_M$ and $\phi_Y$ are dispersion parameters. 

\begin{remark}[Choice of the marginal distribution]
One can also take other parametric, semiparametric, or fully nonparametric ways to model the marginal distributions. Examples include the linear quantile regression \citep{wang2018CopulaBasedQuantileRegression}, the generalized additive model, and kernel/sieve density regression \citep{chen2007Chapter76Large,imai2010GeneralApproachCausal}. The methodology development in this paper can provide a foundation for valuable generalizations.
\end{remark}

\bibliographystyle{biometrika}
\bibliography{QMA_refs_zotero}

\end{document}